\documentclass[10pt]{article}

\usepackage{makeidx}
\usepackage{epsf}
\usepackage{epic}
\usepackage{eepic}
\usepackage{amscd}
\usepackage{amssymb}

\usepackage{citesort}
\usepackage{url}
\usepackage{deleq}
\usepackage{a4}
\usepackage{cite}
\usepackage{ntheorem}

\newcommand{\myqed}{\hfill $\Box$\bigskip}

\renewcommand{\theequation}{\thesection.\arabic{equation}}

\let\ssection=\section

\renewcommand{\section}{\setcounter{equation}{0}\ssection}

\def \Reel{\mathbb{R}}

\def \R {\Reel}

\newcommand{\gK}{generalized Kottler}

\newcommand{\Omegak}{\Omega_k}

\newcommand{\pSo}{\partial M^3} 

\newcommand{\pSi}{\partial_iM^3} 

\newcommand{\mcM}{{\mycal M}}

\newcommand{\bea}{\begin{eqnarray}}
\newcommand{\beaa}{\begin{eqnarray*}}
\newcommand{\bean}{\begin{eqnarray}\nonumber}

\newcommand{\bel}[1]{\begin{equation}\label{#1}}
\newcommand{\beal}[1]{\begin{eqnarray}\label{#1}}
\newcommand{\beadl}[1]{\begin{deqarr}\label{#1}}
\newcommand{\eeadl}[1]{\arrlabel{#1}\end{deqarr}}
\newcommand{\eeal}[1]{\label{#1}\end{eqnarray}}
\newcommand{\eead}[1]{\end{deqarr}}
\newcommand{\eea}{\end{eqnarray}}
\newcommand{\eeaa}{\end{eqnarray*}}

\newcommand{\be}{\begin{equation}}
\newcommand{\ee}{\end{equation}}

\newcommand{\tr}{\mbox{\rm tr}\,}

\newcommand{\eq}[1]{(\ref{#1})}

\newcounter{mnotecount}[section]

\renewcommand{\themnotecount}{\thesection.\arabic{mnotecount}}

\newcommand{\mnote}[1]
{\protect{\stepcounter{mnotecount}}$^{\mbox{\footnotesize $
\bullet$\themnotecount}}$ \marginpar{
\raggedright\tiny\em $\!\!\!\!\!\!\,\bullet$\themnotecount: #1} }

\newcommand{\rmnote}[1]{}
\newcommand{\ptc}[1]{\mnote{{\bf ptc:} #1}}

\newtheorem{Theorem}{Theorem}[section]
\newtheorem{theorem}[Theorem]{Theorem}

\newtheorem{conjecture} [Theorem] {Conjecture}

\DeclareFontFamily{OT1}{rsfs}{}
\DeclareFontShape{OT1}{rsfs}{m}{n}{ <-7> rsfs5 <7-10> rsfs7 <10->
rsfs10}{} \DeclareMathAlphabet{\mycal}{OT1}{rsfs}{m}{n}
\def\scri{{\mycal I}}%
\def\Scri{\scri}

\newcommand{\pMinfty}{\partial_{\infty}M^3}
\newcommand{\pSinfty}{\pMinfty}
\newcommand{\GHn}{Hawking}
\newcommand{\mgh}{M_{Haw}}
\newcommand{\mghu}{\mgh}

\newcommand{\cR}{{\mathcal{R}}}

\newcommand{\real}{{\R}}

\newcommand{\qed}{$\Box$}

\makeindex
\begin{document}

\title{The Penrose Inequality}
\author{             Hubert L. Bray\thanks{
Mathematics Department, 2-179, Massachusetts Institute of
Technology, 77 Massachusetts Avenue, Cambridge, MA  02139, USA,
bray@math.mit.edu. Research supported in part by NSF grants
\#DMS-0206483 and \#DMS-9971960 and by the Erwin Schr\"odinger
Institute, Vienna.}\\ Piotr T. Chru\'{s}ciel\thanks{ D\'epartement
de Math\'ematiques, Facult\'e des Sciences, Parc de Grandmont,
F37200 Tours, France, {Piotr.Chrusciel@lmpt.univ-tours.fr}.
Partially supported by a Polish Research Committee grant 2 P03B
073 24, by the Erwin Schr\"odinger Institute, Vienna, and by a
travel grant from the Vienna City Council; URL \protect\url{
www.phys.univ-tours.fr/}$\sim$\protect\url{piotr}}}
\date{\today} \maketitle

\begin{abstract}
 In 1973, R. Penrose presented an argument
that the total mass of a space-time which contains black holes
with event horizons of total area $A$ should be at least
$\sqrt{A/16\pi}$. An important special case of this physical
statement translates into a very beautiful mathematical inequality
in Riemannian geometry known as the Riemannian Penrose inequality.
This inequality was first established by G. Huisken and T. Ilmanen
in 1997 for a single black hole and then by one of the authors
(HB) in 1999 for any number of black holes. The two approaches use
two different geometric flow techniques and are described here. We
further present some background material concerning the problem at
hand, discuss some applications of Penrose-type inequalities, as
well as the open questions remaining.
\\
\\
AMS Classification Scheme: 83C05, 53C80
\end{abstract}

\tableofcontents

\section{Introduction}
\subsection{What is the Penrose Conjecture?}\label{SPe}

  We will restrict our
attention to statements about space-like slices $(M^3,g,h)$ of a
space-time, where $g$ is the positive definite induced metric on
$M^3$ and $h$ is the second fundamental form of $M^3$ in the
space-time.  From the Einstein equation $G = 8\pi T$, where $G$ is
the Einstein curvature tensor and $T$ is the stress-energy tensor,
it follows from the Gauss and Codazzi equations that
\begin{equation}\label{eqn:c1}
\mu = \frac{1}{8\pi} G^{00} =  \frac{1}{16\pi}
      [R - \sum_{i,j} h^{ij}h_{ij} + (\sum_i h_i{}^i)^2],
\end{equation}
\begin{equation}\label{eqn:c2}
J^i = \frac{1}{8\pi} G^{0i} = \frac{1}{8\pi}\sum_j \nabla_j[h^{ij}
- (\sum_k h_k{}^k)g^{ij}],
\end{equation}
where $\mu$ and $J$ are respectively the energy density and the
current vector density at each point of $M^3$. Then the physical
assumption of nonnegative energy density everywhere in the
space-time as measured by observers moving in all future-pointing,
time-like directions (known as the dominant energy condition)
implies that
\begin{equation}\label{positive}
      \mu \ge |J|
\end{equation}
everywhere on $M^3$.  Hence, we will only consider Cauchy data
$(M^3,g,h)$ which satisfy inequality \eq{positive}.

The final assumption we will make is that $(M^3,g,h)$ is {\it
asymptotically flat}, which will be discussed in more detail
below. Typically, one assumes that $M^3$ consists of a compact set
together with  one or more asymptotically flat ``ends'', each
diffeomorphic to the complement of a ball in $\R^3$. For example,
$\R^3$ has one end, whereas $\R^3 \# \R^3$ has
two ends. 

Penrose's motivation for the Penrose Conjecture~\cite{PenroseSCC}%
\index{Penrose Inequality} goes as follows:  Suppose we begin with
Cauchy data $(M^3,g,h)$ which is asymptotically flat (so that
total mass of a chosen end is defined) and satisfies $\mu \ge |J|$
everywhere. Using this as initial data, solve the Einstein
equation forward in time, and suppose that the resulting
space-time is asymptotically flat in null directions so that the
Trautman-Bondi mass is defined for all retarded times. Suppose
further that the space-time eventually settles down to a Kerr
solution, so that the Trautman-Bondi mass asymptotes to the ADM
mass of the relevant Kerr solution. By the Hawking Area
Theorem~\cite{HE} (compare~\cite{ChDGH}), the total area of the
event horizons of any black holes does not decrease, while the
total Trautman-Bondi mass of the system --- which is expected to
approach the ADM mass at very early advanced times
--- does not increase. Since Kerr solutions all have
\begin{equation}\label{Penrose}
   m \ge \sqrt{A_{e}/16\pi},
\end{equation}
where $m$ is total ADM mass~\cite{dirac:1958,ADM:dynamics} and
$A_e$ is the total area of the event horizons, we must have this
same inequality for the original Cauchy data $(M^3,g,h)$.

The reader will have noticed that the above argument makes a lot
of global assumptions about the resulting space-times, and our
current understanding of the associated mathematical problems is
much too poor to be able to settle those one way or another. The
conjecture that (all, or at least a few key ones of) the above
global properties are satisfied is known under the name of {\em
Penrose's cosmic censorship hypothesis}. We refer the reader to
the article by Lars Andersson in this volume and references
therein for more information about that problem.

 A natural interpretation of the Penrose inequality is
that the mass contributed by a collection of black holes is not
less than $\sqrt{A/16\pi}$. More generally, the question ``How
much matter is in a given region of a space-time?'' is still very
much an open problem~\cite{CY}. In this paper, we will discuss
some of the qualitative aspects of mass in general relativity,
look at examples which are informative, and describe the two very
geometric proofs of the Riemannian Penrose inequality.  The most
general version of the Penrose inequality is still open and is
discussed in section \ref{SPFPC}. The notes here are partly based
on one of the author's (HB) lectures  at the ``Fifty Years of the
Cauchy Problem in General Relativity'' Summer School held in
August 2002 in Carg\`ese (videos of lectures available at URL
\url{http://fanfreluche.math.univ-tours.fr}, or on the DVD
enclosed with this volume), and some sections draw substantially
on his review paper~\cite{Bray-Notices}, following a suggestion of
the editors of this volume. The mathematically oriented reader
with limited knowledge of the associated physics might find it
useful to become acquainted with~\cite{Bray-Notices} before
reading the current presentation.

\subsection{Total Mass in General Relativity}

Amongst the notions of mass which are well understood in general
relativity are local energy density at a point, the total mass of
an asymptotically flat space-time (whether at spacelike or at null
infinity; the former is usually called the ADM mass while the
latter the Trautman-Bondi mass), and the total mass of an
asymptotically anti-de Sitter space-time (often called the
Abbott-Deser mass). On the other hand, defining the mass of a
region larger than a point but smaller than the entire universe is
not very well understood at all. While we will return to this last
question in Section~\ref{SAqlm} below, we start here with a
discussion of the ADM mass.

Suppose $(M^3,g)$ is a Riemannian 3-manifold isometrically
embedded in a (3+1) dimensional Lorentzian space-time $N^4$.
Suppose that $M^3$ has zero second fundamental form in the
space-time.  
(Recall that the second fundamental form is a measure of how much
$M^3$ curves inside $N^4$.  $M^3$ is also sometimes called
``totally geodesic'' since geodesics of $N^4$ which are tangent to
$M^3$ at a point stay inside $M^3$ forever.) The Penrose
inequality (which in its full generality allows for $M^3$ to have
non-vanishing second fundamental form) is known as the
\emph{Riemannian Penrose inequality} when the second fundamental
form is set to zero.\footnote{This terminology is somewhat
misleading, in the following sense: the results discussed below
hold as soon as the scalar curvature is non-negative. This will
certainly be the case if $h_{ij}=0$ and $\mu\ge 0$ in \eq{eqn:c1},
but \emph{e.g.} $\sum_i h_i{}^i=0$, or various other conditions in
this spirit, suffice.}

In this work we will mainly consider $(M^3,g)$ that are
asymptotically flat at infinity, which means that for some compact
set $K$, the ``end'' $M^3 \backslash K$ is diffeomorphic to
$\real^3 \backslash B_1(0)$, where the metric $g$ is
asymptotically approaching (with the decay conditions \eq{afconbd}
below) the standard flat metric $\delta_{ij}$ on $\real^3$ at
infinity. The simplest example of an asymptotically flat manifold
is $(\real^3,\delta_{ij})$ itself. Other good examples are the
conformal metrics $(\real^3, u(x)^4 \delta_{ij})$, where $u(x)$
approaches a constant sufficiently rapidly at infinity. (Also,
sometimes it is convenient to allow $(M^3,g)$ to have multiple
asymptotically flat ends, in which case each connected component
of $M^3 \backslash K$ must have the property described above.) A
qualitative picture of an asymptotically flat 3-manifold is shown
below.

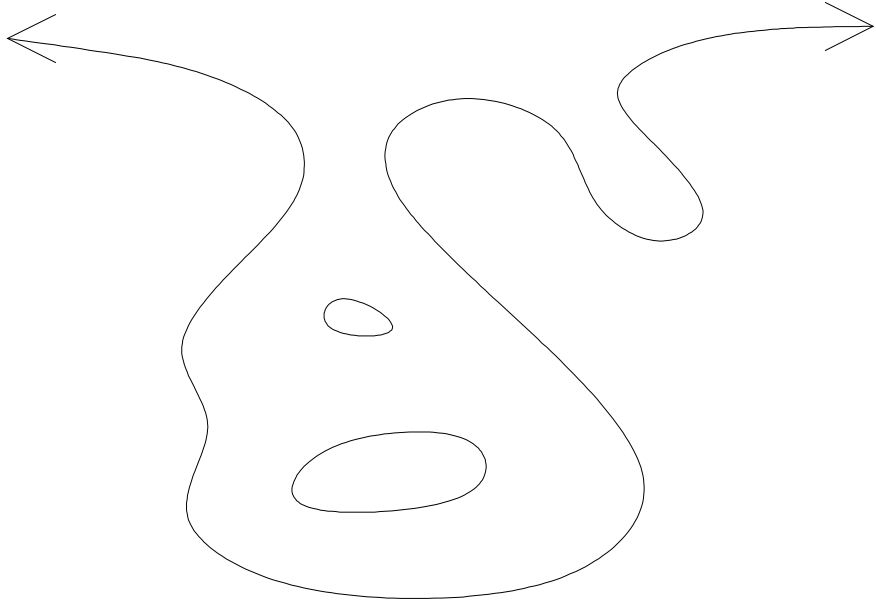
\begin{center}
\setlength{\unitlength}{0.00083333in}
%
{\renewcommand{\dashlinestretch}{30}
\begin{picture}(5424,3760)(0,-10)
\path(312,3658)(12,3508)(312,3358)
\path(5112,3733)(5412,3583)(5112,3433)
\path(12,3508)(14,3508)(19,3507)
	(29,3506)(44,3504)(64,3501)
	(91,3497)(124,3493)(162,3487)
	(206,3481)(254,3475)(305,3468)
	(358,3460)(412,3452)(466,3444)
	(520,3436)(572,3428)(623,3420)
	(672,3413)(718,3405)(762,3398)
	(804,3391)(843,3384)(880,3377)
	(916,3370)(949,3363)(981,3356)
	(1012,3349)(1041,3342)(1070,3335)
	(1097,3328)(1125,3320)(1157,3311)
	(1189,3301)(1221,3291)(1253,3280)
	(1284,3269)(1315,3258)(1346,3246)
	(1377,3233)(1407,3220)(1437,3206)
	(1466,3192)(1494,3178)(1522,3164)
	(1548,3149)(1574,3134)(1599,3119)
	(1622,3104)(1644,3088)(1664,3073)
	(1684,3058)(1702,3043)(1719,3029)
	(1734,3014)(1749,3000)(1762,2985)
	(1775,2970)(1788,2953)(1800,2936)
	(1812,2918)(1822,2899)(1831,2880)
	(1839,2861)(1846,2841)(1852,2821)
	(1857,2800)(1860,2779)(1863,2758)
	(1864,2737)(1864,2716)(1863,2695)
	(1861,2675)(1859,2655)(1855,2636)
	(1850,2617)(1845,2598)(1839,2580)
	(1832,2563)(1825,2545)(1816,2528)
	(1807,2511)(1798,2494)(1787,2476)
	(1775,2458)(1763,2440)(1749,2421)
	(1734,2402)(1719,2382)(1703,2362)
	(1686,2342)(1668,2322)(1650,2301)
	(1631,2281)(1612,2261)(1593,2240)
	(1573,2220)(1554,2200)(1534,2180)
	(1515,2161)(1495,2141)(1475,2120)
	(1456,2102)(1436,2082)(1417,2063)
	(1397,2042)(1376,2021)(1356,1999)
	(1335,1976)(1314,1953)(1293,1930)
	(1272,1905)(1252,1881)(1233,1856)
	(1214,1832)(1196,1807)(1180,1782)
	(1165,1758)(1151,1734)(1139,1711)
	(1128,1688)(1119,1666)(1112,1645)
	(1106,1624)(1102,1603)(1100,1583)
	(1099,1561)(1100,1540)(1104,1518)
	(1108,1496)(1115,1473)(1123,1451)
	(1132,1428)(1142,1404)(1153,1381)
	(1164,1357)(1176,1333)(1188,1309)
	(1200,1285)(1211,1262)(1222,1238)
	(1232,1215)(1241,1193)(1248,1170)
	(1254,1148)(1258,1126)(1261,1105)
	(1262,1083)(1261,1063)(1259,1042)
	(1255,1021)(1250,1000)(1244,977)
	(1237,954)(1229,931)(1220,907)
	(1211,882)(1201,857)(1191,832)
	(1181,806)(1172,781)(1163,755)
	(1155,730)(1147,705)(1141,680)
	(1136,656)(1132,632)(1129,609)
	(1129,586)(1130,564)(1132,542)
	(1137,520)(1142,503)(1149,486)
	(1157,469)(1167,452)(1178,434)
	(1190,417)(1205,399)(1221,382)
	(1238,364)(1258,346)(1279,328)
	(1301,311)(1325,293)(1351,276)
	(1378,259)(1407,243)(1437,227)
	(1467,211)(1499,196)(1532,181)
	(1566,167)(1601,154)(1636,141)
	(1672,129)(1708,118)(1745,107)
	(1783,97)(1821,88)(1860,79)
	(1900,70)(1931,64)(1964,58)
	(1998,53)(2032,48)(2067,43)
	(2103,38)(2139,34)(2177,30)
	(2216,26)(2255,23)(2295,20)
	(2336,18)(2377,16)(2419,14)
	(2462,13)(2505,12)(2548,12)
	(2592,12)(2635,12)(2679,14)
	(2722,15)(2765,17)(2808,19)
	(2851,22)(2892,26)(2933,30)
	(2974,34)(3014,38)(3052,44)
	(3090,49)(3127,55)(3164,61)
	(3199,68)(3233,75)(3266,83)
	(3299,91)(3331,99)(3362,108)
	(3396,118)(3429,129)(3462,141)
	(3494,154)(3525,167)(3556,180)
	(3587,195)(3616,210)(3645,227)
	(3673,244)(3701,261)(3727,280)
	(3753,299)(3778,319)(3801,339)
	(3823,361)(3844,382)(3864,404)
	(3882,427)(3899,450)(3915,473)
	(3928,496)(3941,520)(3951,544)
	(3961,568)(3968,592)(3974,615)
	(3979,639)(3982,663)(3983,687)
	(3983,711)(3982,735)(3979,759)
	(3975,783)(3969,807)(3962,832)
	(3953,857)(3943,883)(3931,909)
	(3918,935)(3904,962)(3888,990)
	(3871,1018)(3852,1047)(3832,1076)
	(3811,1106)(3788,1136)(3765,1166)
	(3740,1197)(3714,1227)(3688,1258)
	(3661,1289)(3633,1319)(3605,1350)
	(3577,1380)(3548,1410)(3519,1440)
	(3490,1469)(3461,1498)(3432,1526)
	(3404,1554)(3375,1581)(3347,1607)
	(3319,1633)(3292,1659)(3265,1684)
	(3238,1709)(3212,1733)(3184,1759)
	(3157,1784)(3129,1810)(3101,1835)
	(3074,1861)(3046,1886)(3019,1912)
	(2991,1938)(2963,1964)(2936,1990)
	(2908,2015)(2881,2041)(2854,2067)
	(2827,2093)(2801,2119)(2775,2144)
	(2750,2169)(2726,2194)(2702,2218)
	(2679,2242)(2657,2265)(2636,2288)
	(2615,2310)(2596,2331)(2578,2352)
	(2560,2373)(2544,2392)(2528,2411)
	(2513,2430)(2500,2448)(2487,2466)
	(2475,2483)(2459,2506)(2445,2528)
	(2433,2550)(2421,2573)(2410,2595)
	(2401,2617)(2392,2639)(2385,2661)
	(2379,2682)(2374,2704)(2371,2725)
	(2368,2746)(2367,2766)(2367,2786)
	(2368,2805)(2371,2824)(2374,2841)
	(2379,2858)(2384,2875)(2391,2890)
	(2398,2905)(2406,2919)(2415,2932)
	(2425,2945)(2436,2959)(2448,2973)
	(2462,2986)(2477,2999)(2493,3012)
	(2511,3024)(2529,3037)(2549,3048)
	(2570,3059)(2592,3070)(2615,3080)
	(2639,3089)(2663,3098)(2688,3105)
	(2713,3112)(2738,3117)(2763,3122)
	(2788,3126)(2812,3129)(2837,3131)
	(2862,3133)(2887,3133)(2910,3133)
	(2933,3132)(2957,3130)(2982,3127)
	(3007,3124)(3032,3120)(3058,3115)
	(3084,3109)(3110,3103)(3137,3095)
	(3163,3087)(3189,3078)(3215,3069)
	(3240,3058)(3264,3048)(3288,3036)
	(3310,3025)(3332,3013)(3352,3000)
	(3371,2987)(3389,2974)(3406,2961)
	(3422,2947)(3437,2933)(3452,2917)
	(3466,2900)(3480,2883)(3493,2865)
	(3505,2846)(3516,2826)(3528,2806)
	(3539,2785)(3549,2763)(3559,2741)
	(3569,2719)(3579,2696)(3588,2674)
	(3598,2652)(3607,2630)(3616,2609)
	(3625,2589)(3635,2569)(3644,2549)
	(3654,2531)(3664,2513)(3675,2495)
	(3687,2477)(3700,2458)(3715,2440)
	(3730,2423)(3747,2405)(3765,2388)
	(3784,2371)(3804,2355)(3824,2340)
	(3846,2325)(3867,2312)(3890,2299)
	(3912,2288)(3934,2278)(3956,2269)
	(3978,2262)(3999,2256)(4021,2251)
	(4041,2248)(4062,2245)(4079,2244)
	(4096,2244)(4114,2245)(4131,2246)
	(4149,2249)(4166,2252)(4184,2256)
	(4201,2261)(4217,2267)(4234,2274)
	(4250,2281)(4264,2290)(4279,2299)
	(4292,2309)(4304,2319)(4314,2330)
	(4324,2342)(4332,2354)(4338,2366)
	(4343,2379)(4347,2392)(4349,2405)
	(4350,2419)(4350,2433)(4347,2449)
	(4343,2466)(4337,2483)(4329,2501)
	(4320,2520)(4309,2540)(4296,2560)
	(4282,2581)(4267,2603)(4250,2625)
	(4232,2647)(4214,2670)(4195,2692)
	(4175,2714)(4155,2736)(4136,2757)
	(4116,2777)(4097,2797)(4078,2817)
	(4060,2835)(4042,2853)(4025,2870)
	(4006,2889)(3988,2908)(3970,2926)
	(3952,2944)(3935,2962)(3919,2981)
	(3903,2999)(3888,3017)(3874,3035)
	(3861,3053)(3850,3070)(3840,3088)
	(3832,3104)(3826,3120)(3821,3136)
	(3818,3151)(3817,3166)(3818,3180)
	(3820,3194)(3825,3208)(3830,3221)
	(3837,3233)(3846,3246)(3856,3259)
	(3868,3272)(3881,3285)(3896,3299)
	(3913,3312)(3932,3326)(3951,3339)
	(3972,3353)(3994,3366)(4017,3379)
	(4040,3391)(4064,3403)(4089,3414)
	(4113,3425)(4138,3435)(4163,3445)
	(4187,3454)(4212,3462)(4237,3470)
	(4260,3477)(4284,3484)(4307,3491)
	(4332,3497)(4357,3503)(4383,3509)
	(4410,3514)(4437,3520)(4465,3525)
	(4493,3530)(4521,3534)(4550,3539)
	(4578,3543)(4606,3546)(4634,3550)
	(4662,3553)(4689,3556)(4716,3559)
	(4742,3561)(4767,3564)(4792,3566)
	(4815,3567)(4839,3569)(4862,3570)
	(4887,3572)(4912,3573)(4937,3575)
	(4962,3576)(4989,3577)(5016,3578)
	(5045,3578)(5076,3579)(5109,3580)
	(5143,3580)(5179,3581)(5216,3581)
	(5252,3582)(5287,3582)(5320,3582)
	(5349,3583)(5373,3583)(5390,3583)
	(5402,3583)(5409,3583)(5412,3583)
\path(2862,658)(2879,667)(2894,676)
	(2909,686)(2922,696)(2934,706)
	(2946,717)(2956,728)(2965,740)
	(2973,752)(2980,764)(2986,777)
	(2990,790)(2994,803)(2996,816)
	(2998,829)(2998,842)(2996,855)
	(2994,868)(2991,881)(2986,893)
	(2981,905)(2974,917)(2967,928)
	(2958,939)(2949,950)(2939,960)
	(2928,969)(2916,978)(2904,986)
	(2891,994)(2877,1001)(2862,1008)
	(2843,1016)(2823,1022)(2801,1028)
	(2778,1034)(2753,1039)(2728,1043)
	(2701,1046)(2673,1049)(2644,1051)
	(2613,1052)(2583,1052)(2551,1052)
	(2519,1051)(2487,1050)(2455,1048)
	(2424,1045)(2392,1042)(2362,1039)
	(2332,1034)(2303,1030)(2274,1025)
	(2247,1020)(2221,1014)(2196,1008)
	(2172,1002)(2150,995)(2121,987)
	(2094,977)(2067,967)(2042,956)
	(2018,945)(1994,933)(1972,920)
	(1950,907)(1930,893)(1911,880)
	(1894,865)(1878,851)(1863,837)
	(1850,823)(1838,809)(1828,796)
	(1819,783)(1812,770)(1805,758)
	(1800,745)(1796,736)(1793,727)
	(1790,718)(1788,709)(1787,700)
	(1786,691)(1787,682)(1788,672)
	(1790,663)(1794,655)(1798,646)
	(1804,637)(1811,629)(1819,621)
	(1828,614)(1839,606)(1851,599)
	(1864,593)(1878,587)(1894,582)
	(1910,577)(1928,572)(1947,568)
	(1967,564)(1989,561)(2012,558)
	(2035,556)(2059,554)(2085,552)
	(2112,551)(2140,550)(2170,550)
	(2201,550)(2234,550)(2268,551)
	(2302,552)(2337,554)(2373,557)
	(2410,559)(2446,563)(2482,567)
	(2518,571)(2554,576)(2588,581)
	(2622,587)(2655,593)(2686,600)
	(2716,607)(2744,614)(2771,622)
	(2796,631)(2820,639)(2842,648)(2862,658)
\path(2412,1695)(2414,1705)(2413,1714)
	(2409,1725)(2403,1736)(2394,1748)
	(2383,1761)(2369,1773)(2353,1786)
	(2335,1799)(2316,1812)(2296,1824)
	(2275,1835)(2253,1846)(2231,1855)
	(2210,1863)(2188,1870)(2168,1876)
	(2148,1880)(2130,1882)(2112,1883)
	(2101,1883)(2090,1882)(2080,1880)
	(2069,1877)(2060,1874)(2051,1870)
	(2042,1866)(2034,1861)(2026,1855)
	(2019,1849)(2012,1842)(2007,1835)
	(2001,1828)(1997,1820)(1993,1812)
	(1990,1803)(1988,1795)(1987,1786)
	(1987,1778)(1987,1769)(1988,1761)
	(1990,1752)(1993,1744)(1997,1736)
	(2002,1729)(2007,1721)(2013,1714)
	(2020,1708)(2028,1701)(2037,1695)
	(2052,1687)(2068,1680)(2087,1673)
	(2107,1667)(2129,1662)(2152,1658)
	(2177,1655)(2201,1652)(2227,1651)
	(2252,1650)(2276,1651)(2299,1652)
	(2321,1655)(2342,1658)(2360,1662)
	(2375,1667)(2389,1673)(2399,1680)
	(2407,1687)(2412,1695)
\end{picture}
}

\end{center}
 \vspace{.1in}

The assumptions on the asymptotic behavior of $(M^3,g)$ at
infinity will be tailored to imply the existence of the limit
\begin{eqnarray}\label{totalmass}
m = \frac{1}{16\pi} \lim_{\sigma\to\infty}
\int_{S_\sigma}\sum_{i,j}(g_{ij,i}\nu_j-g_{ii,j}\nu_j)\,d\mu
\end{eqnarray}
where $S_\sigma$ is the coordinate sphere of radius $\sigma$,
$\nu$ is the unit normal to $S_\sigma$, and $d\mu$ is the area
element of $S_\sigma$ in the coordinate chart. The quantity $m$ is
called the {\em total mass} (or ADM mass~\cite{ADM:dynamics}) of
$(M^3,g)$.\index{ADM mass}\index{mass!ADM} Equation~\eq{totalmass}
begs the question of the geometric character of the number $m$:
the integrand contains partial derivatives of a tensor, which
makes it coordinate dependent. For example, if $g=\delta$ is the
flat metric in the standard orthogonal coordinates $x^i$, one
clearly obtains zero. On the other hand, we can introduce a new
coordinate system $(\rho,\theta,\phi)$ by changing the radial
variable $r$ to
\bel{m48} r= \rho + c\rho^{1-\alpha}\;,\ee
with some constants $\alpha >0$, $c\in \R$. In the associated
asymptotically Euclidean coordinate system $y^i= \rho x^i/r$ the
metric tensor approaches $\delta $ as $O(|y|^{-\alpha})$:
$$\delta_{ij}dx^idx^j= g_{ij}dy^idy^j\;,\
$${with} \bel{afconbd} g_{ij}-\delta_{ij}={\cal O}(|y|^{-\alpha})\;,\ \partial_k
g_{ij} = {\cal O}(|y|^{-\alpha-1})\;.\ee A short calculation gives
\bean m & = &  \cases{ \infty\;, & $\alpha <
1/2\;,$ \cr c^2/8\;, & $\alpha = 1/2\;,$ \cr 0\;,&$\alpha >
1/2\;.$ } \eeal{m50} Thus, the mass $m$ of the flat metric in the
coordinate system $y^i$ is infinite if $\alpha < 1/2$, can have an
arbitrary positive value depending upon $c$ if $\alpha=1/2$, and
vanishes for $\alpha>1/2$. (Negative values of $m$ can also be
obtained by deforming the slice $\{t=0\}$ within Minkowski
space-time~\cite{Chremark} when  the decay rate $\alpha=1/2$ is
allowed.)
 The lesson
of this is that the mass appears to depend upon the coordinate
system chosen, even within the class of coordinate systems in
which the metric tends to a constant coefficients matrix as $r$
tends to infinity. It can be shown that the decay rate
$\alpha=1/2$ is precisely the borderline for a well defined mass:
the mass is an invariant in the class of coordinate systems
satisfying \eq{afconbd} with $\alpha>1/2$ and with $R\in L^1(M)$
\cite{bartnik:mass,ChErice}\footnote{Actually the results
in~\cite{bartnik:mass} use weighted Sobolev conditions on two
derivatives of the metric, suggesting that the right decay
conditions in \eq{afconbd} are $o(|y|^{-1/2})$ for the metric and
$o(|y|^{-3/2})$ for its derivatives. It can be checked that the
argument in~\cite{ChErice} generalises, and gives the result under
those conditions.}. We note that the above example is essentially
due to Denisov and Solov'ev~\cite{DenisovSoloviev83}, and that the
geometric character of $m$ in a space-time setting is established
in~\cite{ChmassCMP}.

Going back to the example $(\real^3, u(x)^4 \delta_{ij})$, if we
suppose that $u(x) > 0$ has the asymptotics at infinity
\begin {equation}\label{eqn:exp}
   u(x) = a + b/|x| + {\cal O}(1/|x|^2)\;,
\end{equation}
with the derivatives of the ${\cal O}(1/|x|^2)$ term being ${\cal
O}(1/|x|^3)$, then the total mass of $(M^3,g)$ is
\begin{equation}\label{eqn:tmass}
   m = 2ab.
\end{equation}
Furthermore, suppose $(M^3,g)$ is any metric whose ``end'' is
isometric to $(\real^3 \backslash K, u(x)^4 \delta_{ij})$, where
$u(x)$ is harmonic in the coordinate chart of the end $(\real^3
\backslash K, \delta_{ij})$ and goes to a constant at infinity.
Then expanding $u(x)$ in terms of spherical harmonics demonstrates
that $u(x)$ satisfies condition \eq{eqn:exp}. We will call these
Riemannian manifolds $(M^3,g)$ {\em harmonically flat at
infinity}, and we note that the total mass of these manifolds is
also given by equation \eq{eqn:tmass}.

A very nice lemma by Schoen and Yau~\cite{SchoenYau81} is that,
given any $\epsilon
>0$, it is always possible to perturb an asymptotically flat
manifold to become harmonically flat at infinity such that the
total mass changes less than $\epsilon$ and the metric changes
less than $\epsilon$ pointwise, all while maintaining nonnegative
scalar curvature (discussed in a moment).  Hence, it happens that
to prove the theorems in this paper, we only need to consider
harmonically flat manifolds. Thus, we can use equation
\eq{eqn:tmass} as our definition of total mass. As an example
(already pointed out), note that $(\real^3,\delta_{ij})$ has zero
total mass. Also, note that, qualitatively, the total mass of an
asymptotically flat or harmonically flat manifold is the $1/r$
rate at which the metric becomes flat at infinity.

A deep (and considerably more difficult to prove) result of
Corvino~\cite{Corvino} (compare~\cite{CorvinoSchoenprep,ChDelay})
shows that if $m$ is non zero, then one can always perturb an
asymptotically flat manifold as above while maintaining zero
scalar curvature and achieving \eq{eqn:exp} without any error
term.

We finish this section by noting the following ``isotropic
coordinates" representation of the {\em exterior Schwarzschild
space-time metric}
\bel{stschw}\left(\real\times \left(\R^3 \setminus B_{m/2}(0)\right),
(1 + \frac{m}{2|x|})^4 (dx_1^2 + dx_2^2 + dx_3^2) -
\left(\frac{1-m/2|x|}{1+m/2|x|}\right)^2 dt^2\right)\;. \ee The
$t=0$ slice (which has zero second fundamental form) is the
\emph{exterior spacelike Schwarzschild metric} \bel{spschw}
\left(\real^3 \backslash B_{m/2}(0),
(1+\frac{m}{2|x|})^4\delta_{ij}\right).\ee According to equation
\eq{eqn:tmass}, the parameter $m$ is of course the total mass of
this $3$-manifold.

The above example also allows us to make a connection between what
we have arbitrarily defined to be total mass and our more
intuitive Newtonian notions of mass.  Using the natural Lorentzian
coordinate chart as a reference, one can compute that geodesics in
the Schwarzschild space-time metric are curved when $m \ne 0$.
Furthermore, if one interprets this curvature as acceleration due
to a force coming from the central region of the manifold, one
finds that this fictitious force yields an acceleration asymptotic
to $m/r^2$ for large $r$.  Hence, a test particle left to drift
along geodesics far out in the asymptotically flat end of the
Schwarzschild spacetime ``accelerates'' according to Newtonian
physics as if the total mass of the system were $m$.

\subsection{Example Using Superharmonic Functions in $\real^3$}

Once again, let us return to the $(\real^3,u(x)^4\delta_{ij})$
example. The formula for the scalar curvature is
\begin{eqnarray*}
   R(x) = -8 u(x)^{-5} \Delta u(x).
\end{eqnarray*}
Hence, since the physical assumption of nonnegative energy density
implies nonnegative scalar curvature, we see that $u(x) > 0$ must
be superharmonic ($\Delta u \le 0$).  For simplicity, let us  also
assume that $u(x)$ is harmonic outside a bounded set so that we
can expand $u(x)$ at infinity using spherical harmonics.  Hence,
$u(x)$ has the asymptotics of equation \eq{eqn:exp}. By the
maximum principle, it follows that the minimum value for $u(x)$
must be $a$, referring to equation \eq{eqn:exp}.  Hence, $b \ge
0$, which implies that $m \ge 0$.  Thus we see that the assumption
of nonnegative energy density at each point of
$(\real^3,u(x)^4\delta_{ij})$ implies that the total mass is also
nonnegative, which is what one would hope.

\subsection{The Positive Mass Theorem}

Suppose we have any asymptotically flat manifold with nonnegative
scalar curvature, is it true that the total mass is also
nonnegative?  The answer is {\it yes}, and this fact is known as
the positive mass theorem, first proved by Schoen and Yau
\cite{SchoenYau79b} in 1979 using minimal surface techniques and
then by Witten~\cite{Witten:mass} in 1981 using spinors. (The
mathematical details needed for Witten's argument have been worked
out
in~\cite{ParkerTaubes82,bartnik:mass,ChBlesHouches,Herzlich:mass}.)
In the zero second fundamental form case, also known as the
\emph{time-symmetric} case, the positive mass theorem is known as
the Riemannian positive mass theorem and is stated below.

\begin{theorem} (Schoen, Yau~\cite{SchoenYau79b})  Let $(M^3,g)$ be any asymptotically flat,
complete Riemannian manifold with nonnegative scalar curvature.
Then the total mass $m \ge 0$, with equality if and only if
$(M^3,g)$ is isometric to $(\real^3,\delta)$.\index{Positive mass
theorem}
\end{theorem}

\subsection{Apparent horizons}

Given a surface in a space-time, suppose that it emits an outward
shell of light.  If the surface area of this shell of light is
decreasing everywhere on the surface, then this is called a
trapped surface.\footnote{The reader is warned that several
authors require the trapping of both outwards and inwards shells
of light in the definition of trapped surface. The inwards null
directions are irrelevant for our purposes, and they are therefore
ignored in the definition here.} The outermost boundary of these
trapped surfaces is called the apparent horizon. Apparent horizons
can be computed in terms of Cauchy data, and under appropriate
global hypotheses an apparent horizon implies the existence of an
event horizon outside of it~\cite{HE,Waldbook} in the
time-symmetric case. The reader is referred
to~\cite{Chrusciel:2002mi} for a review of what is known about
apparent horizons; further recent results
include~\cite{Dain:ah,Maxwell:ah}.

Now let us return to the case where $(M^3,g)$ is a ``$t = 0$''
slice of a space-time with zero second fundamental form.  Then
apparent horizons of black holes intersected with $M^3$ correspond
to the connected components of the outermost minimal surface
$\Sigma_0$ of $(M^3,g)$.

All of the surfaces we are considering in this paper will be
required to be smooth boundaries of open bounded regions, so that
outermost is well-defined with respect to a chosen end of the
manifold~\cite{Bray:preparation2}.  A minimal surface in $(M^3,g)$
is a surface which is a critical point of the area function with
respect to any smooth variation of the surface.  The first
variational calculation implies that minimal surfaces have zero
mean curvature.  The surface $\Sigma_0$ of $(M^3,g)$ is defined as
the boundary of the union of the open regions bounded by all of
the minimal surfaces in $(M^3,g)$.  It turns out that $\Sigma_0$
also has to be a minimal surface, so we call $\Sigma_0$ the {\em
outermost minimal surface}. A qualitative sketch of an outermost
minimal surface of a 3-manifold is shown below.

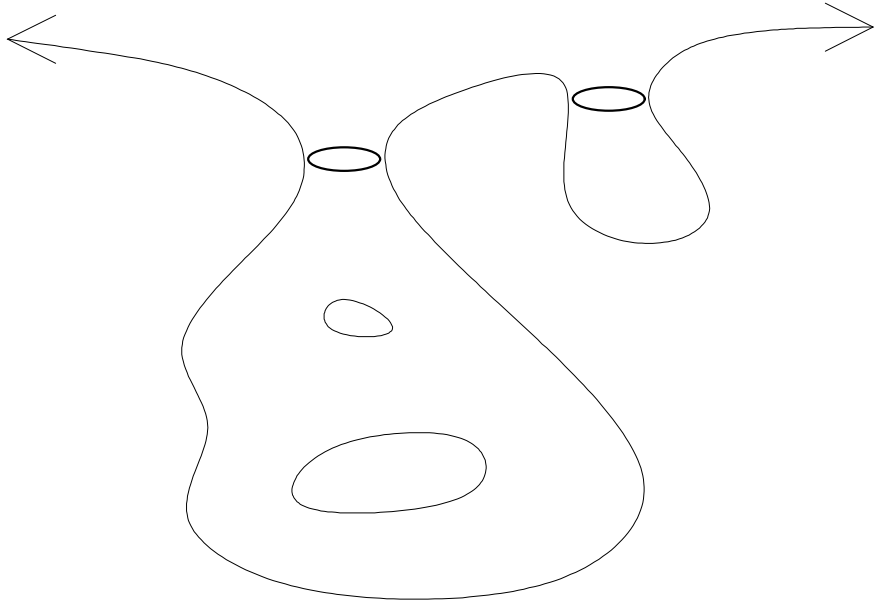
\begin{center}
\setlength{\unitlength}{0.00083333in}
%
{\renewcommand{\dashlinestretch}{30}
\begin{picture}(5424,3760)(0,-10)
\thicklines
\put(2112,2758){\ellipse{450}{150}}
\put(3762,3133){\ellipse{450}{150}}
\thinlines
\path(312,3658)(12,3508)(312,3358)
\path(5112,3733)(5412,3583)(5112,3433)
\path(12,3508)(14,3508)(19,3507)
	(29,3506)(44,3504)(64,3501)
	(91,3497)(124,3493)(162,3487)
	(206,3481)(254,3475)(305,3468)
	(358,3460)(412,3452)(466,3444)
	(520,3436)(572,3428)(623,3420)
	(672,3413)(718,3405)(762,3398)
	(804,3391)(843,3384)(880,3377)
	(916,3370)(949,3363)(981,3356)
	(1012,3349)(1041,3342)(1070,3335)
	(1097,3328)(1125,3320)(1157,3311)
	(1189,3301)(1221,3291)(1253,3280)
	(1284,3269)(1315,3258)(1346,3246)
	(1377,3233)(1407,3220)(1437,3206)
	(1466,3192)(1494,3178)(1522,3164)
	(1548,3149)(1574,3134)(1599,3119)
	(1622,3104)(1644,3088)(1664,3073)
	(1684,3058)(1702,3043)(1719,3029)
	(1734,3014)(1749,3000)(1762,2985)
	(1775,2970)(1788,2953)(1800,2936)
	(1812,2918)(1822,2899)(1831,2880)
	(1839,2861)(1846,2841)(1852,2821)
	(1857,2800)(1860,2779)(1863,2758)
	(1864,2737)(1864,2716)(1863,2695)
	(1861,2675)(1859,2655)(1855,2636)
	(1850,2617)(1845,2598)(1839,2580)
	(1832,2563)(1825,2545)(1816,2528)
	(1807,2511)(1798,2494)(1787,2476)
	(1775,2458)(1763,2440)(1749,2421)
	(1734,2402)(1719,2382)(1703,2362)
	(1686,2342)(1668,2322)(1650,2301)
	(1631,2281)(1612,2261)(1593,2240)
	(1573,2220)(1554,2200)(1534,2180)
	(1515,2161)(1495,2141)(1475,2120)
	(1456,2102)(1436,2082)(1417,2063)
	(1397,2042)(1376,2021)(1356,1999)
	(1335,1976)(1314,1953)(1293,1930)
	(1272,1905)(1252,1881)(1233,1856)
	(1214,1832)(1196,1807)(1180,1782)
	(1165,1758)(1151,1734)(1139,1711)
	(1128,1688)(1119,1666)(1112,1645)
	(1106,1624)(1102,1603)(1100,1583)
	(1099,1561)(1100,1540)(1104,1518)
	(1108,1496)(1115,1473)(1123,1451)
	(1132,1428)(1142,1404)(1153,1381)
	(1164,1357)(1176,1333)(1188,1309)
	(1200,1285)(1211,1262)(1222,1238)
	(1232,1215)(1241,1193)(1248,1170)
	(1254,1148)(1258,1126)(1261,1105)
	(1262,1083)(1261,1063)(1259,1042)
	(1255,1021)(1250,1000)(1244,977)
	(1237,954)(1229,931)(1220,907)
	(1211,882)(1201,857)(1191,832)
	(1181,806)(1172,781)(1163,755)
	(1155,730)(1147,705)(1141,680)
	(1136,656)(1132,632)(1129,609)
	(1129,586)(1130,564)(1132,542)
	(1137,520)(1142,503)(1149,486)
	(1157,469)(1167,452)(1178,434)
	(1190,417)(1205,399)(1221,382)
	(1238,364)(1258,346)(1279,328)
	(1301,311)(1325,293)(1351,276)
	(1378,259)(1407,243)(1437,227)
	(1467,211)(1499,196)(1532,181)
	(1566,167)(1601,154)(1636,141)
	(1672,129)(1708,118)(1745,107)
	(1783,97)(1821,88)(1860,79)
	(1900,70)(1931,64)(1964,58)
	(1998,53)(2032,48)(2067,43)
	(2103,38)(2139,34)(2177,30)
	(2216,26)(2255,23)(2295,20)
	(2336,18)(2377,16)(2419,14)
	(2462,13)(2505,12)(2548,12)
	(2592,12)(2635,12)(2679,14)
	(2722,15)(2765,17)(2808,19)
	(2851,22)(2892,26)(2933,30)
	(2974,34)(3014,38)(3052,44)
	(3090,49)(3127,55)(3164,61)
	(3199,68)(3233,75)(3266,83)
	(3299,91)(3331,99)(3362,108)
	(3396,118)(3429,129)(3462,141)
	(3494,154)(3525,167)(3556,180)
	(3587,195)(3616,210)(3645,227)
	(3673,244)(3701,261)(3727,280)
	(3753,299)(3778,319)(3801,339)
	(3823,361)(3844,382)(3864,404)
	(3882,427)(3899,450)(3915,473)
	(3928,496)(3941,520)(3951,544)
	(3961,568)(3968,592)(3974,615)
	(3979,639)(3982,663)(3983,687)
	(3983,711)(3982,735)(3979,759)
	(3975,783)(3969,807)(3962,832)
	(3953,857)(3943,883)(3931,909)
	(3918,935)(3904,962)(3888,990)
	(3871,1018)(3852,1047)(3832,1076)
	(3811,1106)(3788,1136)(3765,1166)
	(3740,1197)(3714,1227)(3688,1258)
	(3661,1289)(3633,1319)(3605,1350)
	(3577,1380)(3548,1410)(3519,1440)
	(3490,1469)(3461,1498)(3432,1526)
	(3404,1554)(3375,1581)(3347,1607)
	(3319,1633)(3292,1659)(3265,1684)
	(3238,1709)(3212,1733)(3184,1759)
	(3157,1784)(3129,1810)(3101,1835)
	(3074,1861)(3046,1886)(3019,1912)
	(2991,1938)(2963,1964)(2936,1990)
	(2908,2015)(2881,2041)(2854,2067)
	(2827,2093)(2801,2119)(2775,2144)
	(2750,2169)(2726,2194)(2702,2218)
	(2679,2242)(2657,2265)(2636,2288)
	(2615,2310)(2596,2331)(2578,2352)
	(2560,2373)(2544,2392)(2528,2411)
	(2513,2430)(2500,2448)(2487,2466)
	(2475,2483)(2459,2506)(2445,2528)
	(2433,2550)(2421,2573)(2410,2595)
	(2401,2617)(2392,2639)(2385,2661)
	(2379,2682)(2374,2704)(2371,2725)
	(2368,2746)(2367,2766)(2367,2786)
	(2368,2805)(2371,2824)(2374,2841)
	(2379,2858)(2384,2875)(2391,2890)
	(2398,2905)(2406,2919)(2415,2932)
	(2425,2945)(2436,2959)(2448,2973)
	(2462,2986)(2477,3000)(2493,3013)
	(2511,3026)(2530,3039)(2550,3052)
	(2572,3065)(2595,3078)(2619,3090)
	(2643,3103)(2669,3115)(2695,3126)
	(2721,3138)(2748,3149)(2775,3159)
	(2802,3169)(2829,3179)(2857,3189)
	(2884,3199)(2912,3208)(2938,3217)
	(2965,3225)(2992,3234)(3020,3242)
	(3048,3250)(3077,3258)(3107,3265)
	(3136,3272)(3166,3278)(3196,3283)
	(3225,3288)(3254,3291)(3282,3293)
	(3308,3294)(3334,3294)(3358,3292)
	(3381,3288)(3401,3283)(3420,3277)
	(3438,3269)(3453,3260)(3466,3248)
	(3477,3235)(3487,3220)(3494,3205)
	(3500,3188)(3504,3169)(3507,3149)
	(3509,3126)(3510,3102)(3511,3077)
	(3510,3050)(3508,3021)(3506,2992)
	(3504,2961)(3501,2930)(3498,2898)
	(3495,2865)(3492,2833)(3489,2800)
	(3486,2768)(3484,2737)(3483,2706)
	(3482,2676)(3482,2647)(3484,2620)
	(3486,2593)(3489,2568)(3494,2544)
	(3500,2520)(3507,2499)(3516,2478)
	(3526,2457)(3538,2438)(3551,2419)
	(3566,2402)(3582,2385)(3600,2368)
	(3619,2353)(3639,2338)(3661,2325)
	(3683,2312)(3706,2300)(3730,2289)
	(3754,2279)(3779,2271)(3803,2263)
	(3828,2256)(3852,2250)(3876,2245)
	(3900,2241)(3924,2238)(3946,2236)
	(3969,2234)(3991,2233)(4012,2233)
	(4037,2233)(4061,2235)(4085,2237)
	(4109,2240)(4133,2244)(4156,2249)
	(4179,2254)(4201,2261)(4222,2268)
	(4243,2277)(4263,2286)(4281,2296)
	(4299,2307)(4315,2318)(4329,2329)
	(4342,2342)(4353,2354)(4363,2367)
	(4371,2380)(4378,2393)(4383,2407)
	(4387,2420)(4390,2436)(4391,2453)
	(4390,2470)(4388,2488)(4384,2506)
	(4379,2526)(4372,2546)(4364,2567)
	(4355,2588)(4345,2610)(4333,2631)
	(4321,2653)(4308,2674)(4295,2695)
	(4281,2715)(4267,2735)(4253,2754)
	(4240,2773)(4226,2791)(4212,2808)
	(4197,2827)(4181,2846)(4166,2865)
	(4150,2884)(4135,2903)(4119,2922)
	(4104,2942)(4090,2962)(4076,2981)
	(4063,3001)(4051,3020)(4041,3039)
	(4032,3058)(4025,3076)(4019,3094)
	(4015,3111)(4013,3128)(4012,3145)
	(4013,3160)(4014,3174)(4017,3189)
	(4021,3204)(4026,3219)(4033,3235)
	(4041,3251)(4050,3267)(4060,3283)
	(4072,3299)(4085,3315)(4098,3331)
	(4113,3347)(4129,3362)(4145,3376)
	(4162,3390)(4180,3403)(4198,3415)
	(4216,3427)(4235,3438)(4255,3448)
	(4275,3458)(4293,3466)(4313,3474)
	(4334,3482)(4355,3489)(4377,3496)
	(4400,3503)(4424,3509)(4449,3516)
	(4475,3521)(4501,3527)(4528,3532)
	(4555,3537)(4582,3541)(4609,3545)
	(4637,3549)(4664,3553)(4690,3556)
	(4716,3559)(4742,3561)(4767,3564)
	(4792,3566)(4816,3567)(4839,3569)
	(4862,3570)(4887,3572)(4912,3573)
	(4937,3575)(4962,3576)(4989,3577)
	(5016,3578)(5045,3578)(5076,3579)
	(5109,3580)(5143,3580)(5179,3581)
	(5216,3581)(5252,3582)(5287,3582)
	(5320,3582)(5349,3583)(5373,3583)
	(5390,3583)(5402,3583)(5409,3583)(5412,3583)
\path(2862,658)(2879,667)(2894,676)
	(2909,686)(2922,696)(2934,706)
	(2946,717)(2956,728)(2965,740)
	(2973,752)(2980,764)(2986,777)
	(2990,790)(2994,803)(2996,816)
	(2998,829)(2998,842)(2996,855)
	(2994,868)(2991,881)(2986,893)
	(2981,905)(2974,917)(2967,928)
	(2958,939)(2949,950)(2939,960)
	(2928,969)(2916,978)(2904,986)
	(2891,994)(2877,1001)(2862,1008)
	(2843,1016)(2823,1022)(2801,1028)
	(2778,1034)(2753,1039)(2728,1043)
	(2701,1046)(2673,1049)(2644,1051)
	(2613,1052)(2583,1052)(2551,1052)
	(2519,1051)(2487,1050)(2455,1048)
	(2424,1045)(2392,1042)(2362,1039)
	(2332,1034)(2303,1030)(2274,1025)
	(2247,1020)(2221,1014)(2196,1008)
	(2172,1002)(2150,995)(2121,987)
	(2094,977)(2067,967)(2042,956)
	(2018,945)(1994,933)(1972,920)
	(1950,907)(1930,893)(1911,880)
	(1894,865)(1878,851)(1863,837)
	(1850,823)(1838,809)(1828,796)
	(1819,783)(1812,770)(1805,758)
	(1800,745)(1796,736)(1793,727)
	(1790,718)(1788,709)(1787,700)
	(1786,691)(1787,682)(1788,672)
	(1790,663)(1794,655)(1798,646)
	(1804,637)(1811,629)(1819,621)
	(1828,614)(1839,606)(1851,599)
	(1864,593)(1878,587)(1894,582)
	(1910,577)(1928,572)(1947,568)
	(1967,564)(1989,561)(2012,558)
	(2035,556)(2059,554)(2085,552)
	(2112,551)(2140,550)(2170,550)
	(2201,550)(2234,550)(2268,551)
	(2302,552)(2337,554)(2373,557)
	(2410,559)(2446,563)(2482,567)
	(2518,571)(2554,576)(2588,581)
	(2622,587)(2655,593)(2686,600)
	(2716,607)(2744,614)(2771,622)
	(2796,631)(2820,639)(2842,648)(2862,658)
\path(2412,1695)(2414,1705)(2413,1714)
	(2409,1725)(2403,1736)(2394,1748)
	(2383,1761)(2369,1773)(2353,1786)
	(2335,1799)(2316,1812)(2296,1824)
	(2275,1835)(2253,1846)(2231,1855)
	(2210,1863)(2188,1870)(2168,1876)
	(2148,1880)(2130,1882)(2112,1883)
	(2101,1883)(2090,1882)(2080,1880)
	(2069,1877)(2060,1874)(2051,1870)
	(2042,1866)(2034,1861)(2026,1855)
	(2019,1849)(2012,1842)(2007,1835)
	(2001,1828)(1997,1820)(1993,1812)
	(1990,1803)(1988,1795)(1987,1786)
	(1987,1778)(1987,1769)(1988,1761)
	(1990,1752)(1993,1744)(1997,1736)
	(2002,1729)(2007,1721)(2013,1714)
	(2020,1708)(2028,1701)(2037,1695)
	(2052,1687)(2068,1680)(2087,1673)
	(2107,1667)(2129,1662)(2152,1658)
	(2177,1655)(2201,1652)(2227,1651)
	(2252,1650)(2276,1651)(2299,1652)
	(2321,1655)(2342,1658)(2360,1662)
	(2375,1667)(2389,1673)(2399,1680)
	(2407,1687)(2412,1695)
\end{picture}
}

\end{center}
 \vspace{.1in}

We will also define a surface to be {\em (strictly) outer
minimising} if every surface which encloses it has (strictly)
greater area.  Note that outermost minimal surfaces are strictly
outer minimising.  Also, we define a {\em horizon} in our context
to be any minimal surface which is the boundary of a bounded open
region.

It also follows from a stability argument (using the Gauss-Bonnet
theorem interestingly) that each component of an outermost minimal
surface (in a 3-manifold with nonnegative scalar curvature) must
have the topology of a sphere~\cite{MeeksYau}.

Penrose's argument~\cite{PenroseSCC}, presented in
Section~\ref{SPe}, suggests that the mass contributed by the black
holes (thought of as the connected components of $\Sigma_0$)
should be at least $\sqrt{A_0/16\pi}$, where $A_0$ is the area of
$\Sigma_0$. This leads to the following geometric statement:

\vspace{.1in}\noindent {\bf The Riemannian Penrose
Inequality}\index{Riemannian Penrose Inequality} {\it Let
$(M^3,g)$ be a complete, smooth, 3-manifold with nonnegative
scalar curvature which is harmonically flat at infinity with total
mass $m$ and which has an outermost minimal surface $\Sigma_0$ of
area $A_0$. Then
\begin{equation}\label{eqn:RPI}
m \ge \sqrt{\frac{A_0}{16\pi}},
\end{equation}
with equality if and only if $(M^3,g)$ is isometric to the
Schwarzschild metric $(\real^3 \backslash \{0\},
(1+\frac{m}{2|x|})^4\delta_{ij})$ outside their respective
outermost minimal surfaces. }\vspace{.1in}

The above statement has been proved by one of us (HB)
\cite{Bray:preparation2}, and Huisken and Ilmanen~\cite{HI2}
proved it when $A_0$ is defined instead to be the area of the
largest connected component of $\Sigma_0$.   In this paper we will
discuss both approaches, which are very different, although they
both involve flowing surfaces and/or metrics.

We also clarify that the above statement is with respect to a
chosen end of $(M^3,g)$, since both the total mass and the
definition of outermost refer to a particular end.  In fact,
nothing very important is gained by considering manifolds with
more than one end, since extra ends can always be compactified as
follows: Given an extra asymptotically flat end, we can use a
lemma of  Schoen and Yau~\cite{SchoenYau81} to make the end
harmonically flat outside a bounded region. By an extension of
this result in the thesis of one of the authors (HB)
\cite{Bray:thesis}, or using the Corvino-Schoen
construction~\cite{Corvino}, we can make the end exactly
Schwarzschild outside a bounded set while still keeping
nonnegative scalar curvature. We then replace the interior
Schwarzschild region by an object often referred to as ``a bag of
gold", one way of doing it proceeds as follows: Since we are now
in the class of spherically symmetric manifolds, we can then
``round the metric up'' to be an extremely large spherical
cylinder outside a bounded set.  This can be done while keeping
nonnegative scalar curvature since the Hawking mass increases
during this procedure and since the rate of change of the Hawking
mass has the same sign as the scalar curvature in the spherically
symmetric case (as long as the areas of the spheres are
increasing). Finally, the large cylinder can be capped off with a
very large sphere to compactify the end.

Hence, we will typically consider manifolds with just one end. In
the case that the manifold has multiple ends, we will require
every surface (which could have multiple connected components) in
this paper to enclose all of the ends of the manifold except the
chosen end.

\subsection{The Schwarzschild Metric}

The (spacelike) Schwarzschild metric $(\real^3 \backslash \{0\},
(1+\frac{m}{2|x|})^4\delta_{ij})$ (compare~\eq{spschw}), referred
to in the above statement of the Riemannian Penrose Inequality, is
a particularly important example to consider, and corresponds to a
zero-second fundamental form, space-like slice of the usual
(3+1)-dimensional Schwarzschild metric.

The 3-dimensional Schwarzschild metrics    with    total mass $m >
0$ are characterised by being the only spherically symmetric,
geodesically complete, zero scalar curvature 3-metrics, other than
$(\real^3,\delta_{ij})$.  Note that this flat metric on $\real^3$
may be interpreted as the $m=0$ case of the Schwarzschild metric.
Negative values of $m$ also give Schwarzschild metrics, but these
metrics are not geodesically complete since they have a curvature
singularity at the coordinate sphere $r = -m/2$.  If this
singularity is smoothed out in a spherically symmetric way, the
resulting metric has very concentrated negative energy density
(and scalar curvature) in the smoothed out region, which violates
the assumption of positive energy density used throughout this
paper.

The 3-dimensional Schwarzschild metrics with total mass $m > 0$
can also be embedded in 4-dimensional Euclidean space $(x,y,z,w)$
as the set of points satisfying $|(x,y,z)| = \frac{w^2}{8m} + 2m$,
which is a parabola rotated around an $S^2$.  This last picture
allows us to see that the Schwarzschild metric, which has two
ends, has a $Z_2$ symmetry which fixes the sphere with $w=0$ and
$|(x,y,z)| = 2m$, which is clearly minimal.  Furthermore, the area
of this sphere is $4\pi (2m)^2$, giving equality in the Riemannian
Penrose Inequality.

\subsection{A Brief History of the Problem}\label{SHistory}

The Riemannian Penrose Inequality has a rich history spanning
nearly three decades and has motivated much interesting
mathematics and physics. In 1973, R. Penrose in effect conjectured
an even more general version of inequality \eq{eqn:RPI} using a
very clever physical argument~\cite{PenroseSCC}, described in
Section~\ref{SPe}. His observation was that a counterexample to
inequality \eq{eqn:RPI} would yield Cauchy data for solving the
Einstein equations, the solution to which would likely violate the
Cosmic Censor Conjecture (which says that singularities
generically do not form in a space-time unless they are inside a
black hole).

In 1977, Jang and Wald~\cite{JangWald77}, extending ideas of
Geroch~\cite{Geroch:extraction}, gave a heuristic proof of
inequality \eq{eqn:RPI} by defining a flow of 2-surfaces in
$(M^3,g)$ in which the surfaces flow in the outward normal
direction at a rate equal to the inverse of their mean curvatures
at each point.  The Hawking mass of a surface (which is supposed
to estimate the total amount of energy inside the surface) is
defined to be\index{Hawking mass}\index{mass!Hawking}
\begin{eqnarray*}
   m_{Hawking}(\Sigma) = \sqrt{\frac{|\Sigma|}{16\pi}}\left(1-
   \frac{1}{16\pi}\int_\Sigma H^2 \right),
\end{eqnarray*}
(where $|\Sigma|$ is the area of $\Sigma$ and $H$ is the mean
curvature of $\Sigma$ in $(M^3,g)$) and, amazingly, is
nondecreasing under this ``inverse mean curvature flow.'' This is
seen by the fact that under inverse mean curvature flow, it
follows from the Gauss equation and the second variation formula
that
\begin{eqnarray}\label{monoton}
   \frac{d}{dt}m_{Hawking}(\Sigma) = \sqrt{\frac{|\Sigma|}{16\pi}}\left[
   \frac{1}{2} + \frac{1}{16\pi}\int_\Sigma 2\frac{|\nabla_\Sigma H|^2}{H^2}
   + R - 2K + \frac{1}{2}(\lambda_1 - \lambda_2)^2 \right]
\end{eqnarray}
when the flow is smooth, where $R$ is the scalar curvature of
$(M^3,g)$, $K$ is the Gauss curvature of the surface $\Sigma$, and
$\lambda_1$ and $\lambda_2$ are the eigenvalues of the second
fundamental form of $\Sigma$, or principal curvatures.  Hence,
\begin{eqnarray*}
   R \ge 0,
\end{eqnarray*}
and
\begin{equation} \label{eqn:GaussBonnet}
   \int_\Sigma K \le 4\pi
\end{equation}
(which is true for any connected surface by the Gauss-Bonnet
Theorem) imply
\begin{equation}\label{eqn:Hmonotone}
   \frac{d}{dt}m_{Hawking}(\Sigma) \ge 0.
\end{equation}
Furthermore,
\begin{eqnarray*}
   m_{Hawking}(\Sigma_0) = \sqrt{\frac{|\Sigma_0|}{16\pi}}
\end{eqnarray*}
since $\Sigma_0$ is a minimal surface and has zero mean curvature.
In addition, the Hawking mass of sufficiently round spheres at
infinity in the asymptotically flat end of $(M^3,g)$ approaches
the total mass $m$. Hence, if inverse mean curvature flow
beginning with $\Sigma_0$ eventually flows to sufficiently round
spheres at infinity, inequality \eq{eqn:RPI} follows from
inequality \eq{eqn:Hmonotone}.

As noted by Jang and Wald, this argument only works when inverse
mean curvature flow exists and is smooth, which is generally not
expected to be the case. In fact, it is not hard to construct
manifolds which do not admit a smooth inverse mean curvature flow.
One of the main problems is that if the mean curvature of the
evolving surface becomes zero or is negative, it is not clear how
to define the flow.

For twenty years, this heuristic argument lay dormant until the
work of Huisken and Ilmanen~\cite{HI2} in 1997. With a very clever
new approach, Huisken and Ilmanen discovered how to reformulate
inverse mean curvature flow using an energy minimisation principle
in such a way that the new generalised inverse mean curvature flow
always exists.  The added twist is that the surface sometimes
jumps outward.  However, when the flow is smooth, it equals the
original inverse mean curvature flow, and the Hawking mass is
still monotone. Hence, as will be described in the next section,
their new flow produced the first complete proof of inequality
\eq{eqn:RPI} for a single black hole.

Coincidentally, one of the authors (HB) found another proof of
inequality \eq{eqn:RPI}, submitted in 1999, which provides the
correct inequality for any number of black holes. (When the
outermost horizon is not-connected, the Huisken-Ilmanen proof
bounds the mass in terms of the area of its largest component,
while the new argument gives the full inequality, with the sum of
areas of all components.) The approach involves flowing the
original metric to a Schwarzschild metric (outside the horizon) in
such a way that the area of the outermost minimal surface does not
change and the total mass is nonincreasing.  Then since the
Schwarzschild metric gives equality in inequality \eq{eqn:RPI},
the inequality follows for the original metric.

Fortunately, the flow of metrics which is defined is relatively
simple, and in fact stays inside the conformal class of the
original metric.  The outermost minimal surface flows outward in
this conformal flow of metrics, and encloses any compact set (and
hence all of the topology of the original metric) in a finite
amount of time.  Furthermore, this conformal flow of metrics
preserves nonnegative scalar curvature.  We will describe this
approach later in the paper.

Other contributions to the Penrose Conjecture have  been made by
O'Murchadha and Malec in spherical symmetry~\cite{MalecOM94}, by
Herzlich~\cite{Herzlich:Grenoble,Herzlich:inequality} using the
Dirac operator with spectral boundary conditions
(compare~\cite{BartnikChrusciel1,Malec:1998wx}), by Gibbons in the
special case of collapsing shells~\cite{Gibbons:1997js}, by
Tod~\cite{Tod:hoop} as it relates to the hoop conjecture, by
Bartnik~\cite{Bartnik89} for quasi-spherical metrics, by
Jezierski~\cite{Jezierski89,JacekPenroseAPP} using adapted
foliations, and by one of the authors (HB) using isoperimetric
surfaces~\cite{Bray:thesis}. A proof of the Penrose inequality for
conformally flat manifolds (but with suboptimal constant)
 has been given in~\cite{BrayIga}. We also
mention work of Ludvigsen and Vickers~\cite{LudvigsenVickers83b}
using spinors and Bergqvist~\cite{Bergqvist97a}, both concerning
the Penrose inequality for null slices of a space-time.

Various space-time flows which could be used to prove the full
Penrose inequality (see Section~\ref{SPFPC} below) have been
proposed by Hayward~\cite{Hayward:Penrose}, by Mars, Malec and
Simon~\cite{MMS}, and by Frauendiener~\cite{Frauendiener:Penrose}.
It was independently observed by several researchers (HB, Hayward,
Mars, Simon) that those are special cases of the same flow, namely
flowing in the direction $\vec{I} + c(t)\vec{I}'$, where $\vec{I}$
is the inverse mean curvature vector $-\vec{H}/<\vec H,\vec H>$
(which is required to be spacelike outward pointing), $\vec{I}'$
is the future pointing vector with the same length as $\vec{I}$
and orthogonal to $\vec{I}$ in the normal bundle to the surface,
and $\vec{H}$ is the mean curvature vector of the surface in the
spacetime.  The function $c(t)$ is required to satisfy $-1 \le
c(t) \le 1$ but is otherwise free, with its endpoint values
corresponding to Hayward's null flows, $c(t) = 0$ corresponding to
Frauendiener's flow, and $-1 \le c(t) \le 1$ yielding
hypersurfaces satisfying the Mars, Malec, Simon condition which
implies the monotonicity of the spacetime Hawking mass functional.
The catch, however, is that this flow is not parabolic and
therefore only exists for a positive amount of time under special
circumstances. However, as observed by HB at the Penrose
Ineqalities Workshop in Vienna, July 2003, there does exist a way
of defining what a weak solution to the above flow is using a
max-min method analogous to the notion of weak solution to inverse
mean curvature flow (which minimizes an energy functional) defined
by Huisken and Ilmanen \cite{HI2}. Finding ways of constructing
solutions which exist for an infinite amount of time (analogous to
the time-symmetric inverse mean curvature flow due to Huisken and
Ilmanen) is a very interesting problem to consider.

\section{Inverse Mean Curvature Flow}\index{Inverse mean curvature
flow}
 Geometrically, Huisken and Ilmanen's idea
can be described as follows. Let $\Sigma(t)$ be the surface
resulting from inverse mean curvature flow for time $t$ beginning
with the minimal surface $\Sigma_0$. Define $\bar\Sigma(t)$ to be
the outermost minimal area enclosure of $\Sigma(t)$.  Typically,
$\Sigma(t) = \bar\Sigma(t)$ in the flow, but in the case that the
two surfaces are not equal, immediately replace $\Sigma(t)$ with
$\bar\Sigma(t)$ and then continue flowing by inverse mean
curvature.

An immediate consequence of this modified flow is that the mean
curvature of $\bar\Sigma(t)$ is always nonnegative by the first
variation formula, since otherwise $\bar\Sigma(t)$ would be
enclosed by a surface with less area. This is because if we flow a
surface $\Sigma$ in the outward direction with speed $\eta$, the
first variation of the area is $\int_{\Sigma} H\eta$, where $H$ is
the mean curvature of $\Sigma$.

Furthermore, by stability, it follows that in the regions where
$\bar\Sigma(t)$ has zero mean curvature, it is always possible to
flow the surface out slightly to have positive mean curvature,
allowing inverse mean curvature flow to be defined, at least
heuristically at this point.

It turns out that the Hawking mass is still monotone under this
new modified flow. Notice that when $\Sigma(t)$ jumps outward to
$\bar\Sigma(t)$,
\begin{eqnarray*}
\int_{\bar\Sigma(t)} H^2 \le \int_{\Sigma(t)} H^2
\end{eqnarray*}
since $\bar\Sigma(t)$ has zero mean curvature where the two
surfaces do not touch.  Furthermore,
\begin{eqnarray*}
|\bar\Sigma(t)| = |\Sigma(t)|
\end{eqnarray*}
since (this is a neat argument) $|\bar\Sigma(t)| \le |\Sigma(t)|$
(since $\bar\Sigma(t)$ is a minimal area enclosure of $\Sigma(t)$)
and we can not have $|\bar\Sigma(t)| < |\Sigma(t)|$ since
$\Sigma(t)$ would have jumped outward at some earlier time.  This
is only a heuristic argument, but we can then see that the Hawking
mass is nondecreasing during a jump by the above two equations.

This new flow can be rigorously defined, always exists, and the
Hawking mass is monotone, if the scalar curvature is positive.
In~\cite{HI2}, Huisken and Ilmanen define $\Sigma(t)$ to be the
level sets of a scalar valued function $u(x)$ defined on $(M^3,g)$
such that $u(x) = 0$ on the original surface $\Sigma_0$ and
satisfies
\begin{equation}\label{eqn:weak}
\mbox{div}\left(\frac{\nabla u}{|\nabla u|}\right) = |\nabla u|
\end{equation}
in an appropriate weak sense.  Since the left hand side of the
above equation is the mean curvature of the level sets of $u(x)$
and the right hand side is the reciprocal of the flow rate, the
above equation implies inverse mean curvature flow for the level
sets of $u(x)$ when $|\nabla u(x)| \ne 0$.

Huisken and Ilmanen use an energy minimisation principle to define
weak solutions to equation \eq{eqn:weak}. Equation \eq{eqn:weak}
is said to be weakly satisfied in $\Omega$ by the locally
Lipschitz function $u$ if for all locally Lipschitz $v$ with $\{v
\ne u\} \subset\subset \Omega$,
\begin{eqnarray*}
   J_u(u) \le J_u(v)
\end{eqnarray*}
where
\begin{eqnarray*}
   J_u(v) := \int_\Omega |\nabla v| + v|\nabla u|.
\end{eqnarray*}
It can then be seen that the Euler-Lagrange equation of the above
energy functional yields equation \eq{eqn:weak}.

In order to prove that a solution $u$ exists to the above two
equations, Huisken and Ilmanen regularise the degenerate elliptic
equation \eq{eqn:weak} to the elliptic equation
\begin{eqnarray*}
\mbox{div}\left(\frac{\nabla u}{\sqrt{|\nabla u|^2 +
\epsilon^2}}\right) = \sqrt{|\nabla u|^2 + \epsilon^2}.
\end{eqnarray*}
Solutions to the above equation are then shown to exist using the
existence of a subsolution, and then taking the limit as
$\epsilon$ goes to zero yields a weak solution to equation
\eq{eqn:weak}.  There are many details which we are skipping here,
but these are the main ideas.

As it turns out, weak solutions $u(x)$ to equation \eq{eqn:weak}
often have flat regions where $u(x)$ equals a constant.  Hence,
the levels sets $\Sigma(t)$ of $u(x)$ will be discontinuous in $t$
in this case, which corresponds to the ``jumping out'' phenomenon
referred to at the beginning of this section.

We also note that since the Hawking mass of the levels sets of
$u(x)$ is monotone, this inverse mean curvature flow technique not
only proves the Riemannian Penrose Inequality, but also gives a
new proof of the Positive Mass Theorem in dimension three.  This
is seen by letting the initial surface be a very small, round
sphere (which will have approximately zero Hawking mass) and then
flowing by inverse mean curvature, thereby proving $m \ge 0$.

The Huisken and Ilmanen inverse mean curvature flow also seems
ideally suited for proving Penrose inequalities for 3-manifolds
which have $R \ge -6$ and which are asymptotically hyperbolic;
this is discussed in more detail in Section~\ref{Sah}.

Because the monotonicity of the Hawking mass relies on the
Gauss-Bonnet theorem, these arguments do not work in higher
dimensions, at least so far.  Also, because of the need for
equation \eq{eqn:GaussBonnet}, inverse mean curvature flow only
proves the Riemannian Penrose Inequality for a single black hole.
In the next section, we present a technique which proves the
Riemannian Penrose Inequality for any number of black holes, and
which can likely be generalised to higher dimensions.

\section{The Conformal Flow of Metrics}\label{sec:cfm}

Given any initial Riemannian manifold $(M^3,g_0)$ which has
nonnegative scalar curvature and which is harmonically flat at
infinity, we will define a continuous, one parameter family of
metrics $(M^3,g_t)$, $0 \le t < \infty$. This family of metrics
will converge to a 3-dimensional Schwarzschild metric and will
have other special properties which will allow us to prove the
Riemannian Penrose Inequality for the original metric $(M^3,g_0)$.

In particular, let $\Sigma_0$ be the outermost minimal surface of
$(M^3,g_0)$ with area $A_0$.  Then we will also define a family of
surfaces $\Sigma(t)$ with $\Sigma(0) = \Sigma_0$ such that
$\Sigma(t)$ is minimal in $(M^3,g_t)$.  This is natural since as
the metric $g_t$ changes, we expect that the location of the
horizon $\Sigma(t)$ will also change. Then the interesting
quantities to keep track of in this flow are $A(t)$, the total
area of the horizon $\Sigma(t)$ in $(M^3,g_t)$, and $m(t)$, the
total mass of $(M^3,g_t)$ in the chosen end.

In addition to all of the metrics $g_t$ having nonnegative scalar
curvature, we will also have the very nice properties that
\begin{eqnarray*}
A'(t) & = &   0 ,  \\
m'(t) & \le & 0
\end{eqnarray*}
for all $t \ge 0$.  Then since $(M^3,g_t)$ converges to a
Schwarzschild metric (in an appropriate sense) which gives
equality in the Riemannian Penrose Inequality as described in the
introduction,
\begin{equation}\label{eqn:argument}
m(0) \ge m(\infty) = \sqrt{\frac{A(\infty)}{16\pi}}
     = \sqrt{\frac{A(0)}{16\pi}}
\end{equation}
which proves the Riemannian Penrose Inequality for the original
metric $(M^3,g_0)$.  The hard part, then, is to find a flow of
metrics which preserves nonnegative scalar curvature and the area
of the horizon, decreases total mass, and converges to a
Schwarzschild metric as $t$ goes to infinity. This proceeds as
follows:

The metrics $g_t$ will all be conformal to $g_0$.  This conformal
flow of metrics can be thought of as the solution to a first order
o.d.e.~in $t$ defined by equations \eq{eqn:ODE1}-\eq{eqn:ODE4}.
Let
\begin{equation}\label{eqn:ODE1}
   g_t = u_t(x)^4 g_0
\end{equation}
and $u_0(x) \equiv 1$.  Given the metric $g_t$, define
\begin{equation}\label{eqn:ODE2}
   \Sigma(t) = \mbox{the outermost minimal area enclosure of }
               \Sigma_0 \mbox{ in } (M^3,g_t)
\end{equation}
where $\Sigma_0$ is the original outer minimising horizon in
$(M^3,g_0)$. In the cases in which we are interested, $\Sigma(t)$
will not touch $\Sigma_0$, from which it follows that $\Sigma(t)$
is actually a strictly outer minimising horizon of $(M^3,g_t)$.
Then given the horizon $\Sigma(t)$, define $v_t(x)$ such that
\begin{equation}\label{eqn:ODE3}
\left\{
\begin{array}{r l l l}
\Delta_{g_0} v_t(x) & \equiv & 0 & \mbox{ outside } \Sigma(t) \\
v_t(x) & = & 0 & \mbox{ on } \Sigma(t) \\
\lim_{x \rightarrow \infty} v_t(x) & = & -e^{-t} & \\
\end{array}
\right.
\end{equation}
and $v_t(x) \equiv 0$ inside $\Sigma(t)$.  Finally, given
$v_t(x)$, define
\begin{equation}\label{eqn:ODE4}
u_t(x) = 1 + \int_0^t v_s(x) ds
\end{equation}
so that $u_t(x)$ is continuous in $t$ and has $u_0(x) \equiv 1$.

Note that equation \eq{eqn:ODE4} implies that the first order rate
of change of $u_t(x)$ is given by $v_t(x)$.  Hence, the first
order rate of change of $g_t$ is a function of itself, $g_0$, and
$v_t(x)$ which is a function of $g_0$, $t$, and $\Sigma(t)$ which
is in turn a function of $g_t$ and $\Sigma_0$.  Thus, the first
order rate of change of $g_t$ is a function of $t$, $g_t$, $g_0$,
and $\Sigma_0$. (All the results in this section are
from~\cite{Bray:preparation2}.)

\begin{theorem}\label{thm:existence}
Taken together, equations \eq{eqn:ODE1}-\eq{eqn:ODE4} define a
first order o.d.e.~in $t$ for $u_t(x)$ which has a solution which
is Lipschitz in the $t$ variable, $C^1$ in the $x$ variable
everywhere, and smooth in the $x$ variable outside $\Sigma(t)$.
Furthermore, $\Sigma(t)$ is a smooth, strictly outer minimising
horizon in $(M^3,g_t)$ for all $t \ge 0$, and $\Sigma(t_2)$
encloses but does not touch $\Sigma(t_1)$ for all $t_2 > t_1 \ge
0$.
\end{theorem}

Since $v_t(x)$ is a superharmonic function in $(M^3,g_0)$
(harmonic everywhere except on $\Sigma(t)$, where it is weakly
superharmonic), it follows that $u_t(x)$ is superharmonic as well.
Thus, from equation \eq{eqn:ODE4} we see that $\lim_{x \rightarrow
\infty} u_t(x) = e^{-t}$ and consequently that $u_t(x) > 0$ for
all $t$ by the maximum principle.  Then since
\begin{equation}\label{eqn:sc3}
R(g_t) = u_t(x)^{-5}(-8 \Delta_{g_0} + R(g_0))u_t(x)
\end{equation}
it follows that $(M^3,g_t)$ is an asymptotically flat manifold
with nonnegative scalar curvature.

Even so, it still may not seem like $g_t$ is particularly
naturally defined since the rate of change of $g_t$ appears to
depend on $t$ and the original metric $g_0$ in equation
\eq{eqn:ODE3}.  We would prefer a flow where the rate of change of
$g_t$ can be defined purely as a function of $g_t$ (and $\Sigma_0$
perhaps), and interestingly enough this actually does turn out to
be the case. In~\cite{Bray:preparation2} we prove this very
important fact and define a new equivalence class of metrics
called the harmonic conformal class. Then once we decide to find a
flow of metrics which stays inside the harmonic conformal class of
the original metric (outside the horizon) and keeps the area of
the horizon $\Sigma(t)$ constant, then we are basically forced to
choose the particular conformal flow of metrics defined above.

\begin{theorem}\label{thm:monotone}
The function $A(t)$ is constant in $t$ and $m(t)$ is
non-increasing in $t$, for all $t \ge 0$.
\end{theorem}

The fact that $A'(t) = 0$ follows from the fact that to first
order the metric is not changing on $\Sigma(t)$ (since $v_t(x) =
0$ there) and from the fact that to first order the area of
$\Sigma(t)$ does not change as it moves outward since $\Sigma(t)$
is a critical point for area in $(M^3,g_t)$. Hence, the
interesting part of theorem \eq{thm:monotone} is proving that
$m'(t) \le 0$. Curiously, this follows from a nice trick using the
Riemannian positive mass theorem.

Another important aspect of this conformal flow of the metric is
that outside the horizon $\Sigma(t)$, the manifold $(M^3,g_t)$
becomes more and more spherically symmetric and ``approaches'' a
Schwarzschild manifold $(\real^3 \backslash \{0\}, s)$ in the
limit as $t$ goes to $\infty$.  More precisely,

\begin{theorem}\label{thm:limit}
For sufficiently large $t$, there exists a diffeomorphism $\phi_t$
between $(M^3,g_t)$ outside the horizon $\Sigma(t)$ and a fixed
Schwarzschild manifold $(\real^3 \backslash \{0\}, s)$ outside its
horizon.  Furthermore, for all $\epsilon > 0$, there exists a $T$
such that for all $t>T$, the metrics $g_t$ and $\phi^{*}_t(s)$
(when determining the lengths of unit vectors of $(M^3,g_t)$) are
within $\epsilon$ of each other and the total masses of the two
manifolds are within $\epsilon$ of each other.  Hence,
\begin{eqnarray*}
\lim_{t \rightarrow \infty} \frac{m(t)}{\sqrt{A(t)}} =
\sqrt{\frac{1}{16\pi}}.
\end{eqnarray*}
\end{theorem}

Theorem \ref{thm:limit} is not that surprising really although a
careful proof is reasonably long.  However, if one is willing to
believe that the flow of metrics converges to a spherically
symmetric metric outside the horizon, then theorem \ref{thm:limit}
follows from two facts.  The first fact is that the scalar
curvature of $(M^3,g_t)$ eventually becomes identically zero
outside the horizon $\Sigma(t)$ (assuming $(M^3,g_0)$ is
harmonically flat). This follows from the facts that $\Sigma(t)$
encloses any compact set in a finite amount of time, that
harmonically flat manifolds have zero scalar curvature outside a
compact set, that $u_t(x)$ is harmonic outside $\Sigma(t)$, and
equation \eq{eqn:sc3}.  The second fact is that the Schwarzschild
metrics are the only complete, spherically symmetric 3-manifolds
with zero scalar curvature (except for the flat metric on $R^3$).

The Riemannian Penrose inequality, inequality \eq{eqn:RPI}, then
follows from equation \eq{eqn:argument} using theorems
\ref{thm:existence}, \ref{thm:monotone} and \ref{thm:limit}, for
harmonically flat manifolds~\cite{Bray:preparation2}. Since
asymptotically flat manifolds can be approximated arbitrarily well
by harmonically flat manifolds while changing the relevant
quantities arbitrarily little, the asymptotically flat case also
follows. Finally, the case of equality of the Penrose inequality
follows from a more careful analysis of these same arguments.

We refer the reader to~\cite{BraySchoen,BrayICM,Bray-Notices} for
further review-type discussions of the results described above.

\section{Open Questions and Applications}



Now that the Riemannian Penrose conjecture has been proved, what
are the next interesting directions?  What applications can be
found?  Is this subject only of physical interest, or are there
possibly broader applications to other problems in mathematics?

Clearly the most natural open problem is to find a way to prove
the general Penrose conjecture (discussed in the next subsection)
in which $M^3$ is allowed to have any second fundamental form in
the space-time.  There is good reason to think that this may
follow from the Riemannian Penrose inequality, although this is a
bit delicate.  On the other hand, the general positive mass
theorem followed from the Riemannian positive mass theorem as was
originally shown by Schoen and Yau using an idea due to Jang
\cite{SchoenYauPRL,SchoenYau81}. For physicists this problem is
definitely a top priority since most space-times do not even admit
zero second fundamental form space-like slices. We note that the
Riemannian Penrose inequality does give a result which applies to
situations more general than time symmetric, as the condition
$R\ge 0$ holds, \emph{e.g.}, for maximal initial data sets
$\tr\!_g h=0$, as well as in several other situations (``polar
gauge", and so on). However, the general situation remains
open.

 Another interesting question is to ask
these same questions in higher dimensions. One of us (HB) is
currently working on a paper to prove the Riemannian Penrose
inequality in dimensions less than 8. Dimension 8 and higher are
harder because of the surprising fact that minimal hypersurfaces
(and hence apparent horizons of black holes) can have codimension
7 singularities (points where the hypersurface is not smooth).
This curious technicality is also the reason that the positive
mass theorem in dimensions 8 and higher for manifolds which are
not spin has only been announced very recently by Christ and
Lohkamp~\cite{ChristLohkamp}, using a formidable singularity
excision argument,  and it is conceivable that this technique will
allow one to extend the Riemannian Penrose Inequality proof to all
dimensions.

Naturally it is harder to tell what the applications of these
techniques might be to other problems, but already there have been
some.  One application is to the famous Yamabe problem:  Given a
compact 3-manifold $M^3$, define $E(g) = \int_{M^3} R_g dV_g$
where $g$ is scaled so that the total volume of $(M^3,g)$ is one,
$R_g$ is the scalar curvature at each point, and $dV_g$ is the
volume form.  An idea due to Yamabe was to try to construct
canonical metrics on $M^3$ by finding critical points of this
energy functional on the space of metrics.  Define $C(g)$ to be
the infimum of $E(\bar{g})$ over all metrics $\bar{g}$ conformal
to $g$.  Then the (smooth) Yamabe invariant of $M^3$, denoted here
as $Y(M^3)$, is defined to be the supremum of $C(g)$ over all
metrics $g$. $Y(S^3) = 6 \cdot (2\pi^2)^{2/3} \equiv \ Y_1$ is
known to be the largest possible value for Yamabe invariants of
3-manifolds.  It is also known that $Y(T^3) = 0$ and $Y(S^2 \times
S^1) = Y_1 = Y(S^2 \tilde\times S^1)$, where $S^2 \tilde\times
S^1$ is the non-orientable $S^2$ bundle over $S^1$.

One of the authors (HB) and Andre Neves, working on a problem
suggested by Richard Schoen, were able to compute the Yamabe
invariant of $RP^3$\index{Yamabe invariant of $RP^3$} using
inverse mean curvature flow techniques \cite{Bray_Neves} (see
also~\cite[Lecture~2]{BrayCargese}) and found that $Y(RP^3) = Y_1
/ 2^{2/3} \equiv Y_2$.  A corollary is $Y(RP^2 \times S^1) = Y_2$
as well. These techniques also yield the surprisingly strong
result that the only prime 3-manifolds with Yamabe invariant
larger than $RP^3$ are $S^3$, $S^2 \times S^1$, and $S^2
\tilde\times S^1$. The Poincar\'e conjecture for 3-manifolds with
Yamabe invariant greater than $RP^3$ is therefore a corollary.
Furthermore, the problem of classifying 3-manifolds is known to
reduce to the problem of classifying prime 3-manifolds.  The
Yamabe approach then would be to make a list of prime 3-manifolds
ordered by Y.  The first five prime 3-manifolds on this list are
therefore $S^3$, $S^2 \times S^1$, $S^2 \tilde\times S^1$, $RP^3$,
and $R{P}^2 \times S^1$.

\subsection{The Riemannian Penrose conjecture on asymptotically
hyperbolic manifolds}\label{Sah}

 Another natural class of
metrics that are of interest in general relativity consists of
metrics which asymptote to the hyperbolic metric. Such metrics
arise when considering solutions with a negative cosmological
constant, or when considering ``hyperboloidal hypersurfaces" in
space-times which are asymptotically flat in isotropic directions
(technically speaking, these are spacelike hypersurfaces which
intersect $\scri$ transversally in the conformally completed
space-time). For instance, recall that in the presence of a
cosmological constant $\Lambda$ the scalar constraint equation
reads
$$R= 16\pi \mu + |h|_g^2-(\tr_g h ) ^2 + 2 \Lambda\;.$$
Suppose that $h=\lambda g$, where $\lambda$ is a constant; such an
$h$ solves the vector constraint equation. We then have
\bel{Thetadef}R= 16\pi \mu -6\lambda^2 + 2 \Lambda=:16\pi \mu
+2\Theta\;.\ee The constant $\Theta$ equals thus $\Lambda$ when
$\lambda=0$, or $-3 \lambda^2$ when $\Lambda=0$. The positive
energy condition $\mu\ge0$ is now equivalent to $$R\ge 2
\Theta\;.$$ For $\lambda=0$ the associated model space-time
metrics take the form
\be
  ds^2  =  -(k - \frac {2m}r - \frac \Lambda 3 r^2) dt^2 + (k - \frac
  {2m}r - \frac \Lambda 3 r^2)^{-1} dr^2 + r^2 d\Omegak^2\ , \quad k =
  0, \pm 1\;, \label{Kot}
\ee where $d\Omegak^2$ denotes a metric of constant Gauss
curvature $k$ on a two dimensional compact manifold $M^2$. These
are well known static solutions of the vacuum Einstein equation
with a cosmological constant $\Lambda$; some subclasses of
\eq{Kot}  have been discovered by de Sitter \cite{deSitter1917b}
(\eq{Kot} with $m = 0$ and $k=1$), by Kottler \cite{Kottler}
(Equation~\eq{Kot} with an arbitrary $m$ and $k=1$).  The
parameter $m\in \R$ can be seen to be proportional to the total
\GHn\ mass ({\em cf.}\/~\eq{dHmn} below) of the foliation
$t=\mbox{const}$, $r=\mbox{const}$. We will refer to those
solutions as the \gK\ solutions.  The constant $\Lambda$ in
\eq{Kot} is an arbitrary real number, but in this section we will
only consider $\Lambda <0$.

{}From now on the overall approach resembles closely that for
asymptotically flat space-times, as described earlier in this
work. For instance, one considers manifolds which contain
asymptotic ends diffeomorphic to $\R^+\times M^2$. It is
convenient to think of each of the sets ``$\{r=\infty\}\times
M^2$" as a connected component at infinity of a boundary at
infinity, call it $\pMinfty$,  of the initial data surface $M^3$.
There is a well defined notion of mass for metrics which asymptote
to the above model metrics in the asymptotic ends, somewhat
similar to that in \eq{totalmass}. In the hyperbolic case the
boundary conditions are considerably more delicate to formulate as
compared to the asymptotically flat one, and we refer the reader
to~\cite{ChNagyATMP,Wang,CJL,ChHerzlich} for details. In the case
when $M^3$ arises from a space-times with negative cosmological
constant $\Lambda$, the resulting mass is usually called the
Abbott-Deser mass~\cite{AbbottDeser}\index{Abbott-Deser
mass}\index{mass!Abbott-Deser}; when $\Lambda=0$ and $M^3$ is a
hyperboloidal hypersurface the associated mass is called the
Trautman-Bondi mass\index{Trautman-Bondi
mass}\index{mass!Trautman-Bondi}. (The latter notion of mass has
often been referred to as ``Bondi mass" in the literature, but the
name ``Trautman-Bondi mass" seems more appropriate, in view of the
work in~\cite{T}, which precedes~\cite{BBM} by four years; see
also~\cite{Tlectures}.) A large class of initial data sets with
the desired asymptotic behavior has been constructed
in~\cite{AndChDiss,ACF,Kannar:adS}, and the existence of the
associated space-times has been established
in~\cite{Friedrich:aDS,friedrich:cauchy}.

The monotonicity argument of  Geroch~\cite{Geroch:extraction},
described in Section~\ref{SHistory}, has been extended by Gibbons
\cite{GibbonsGPI}  to accommodate for the negative cosmological
constant; we follow the presentation in~\cite{ChruscielSimon}: We
assume that we are given a three dimensional manifold ${(M^3,g)}$
with
 connected minimal boundary $\partial M^3$ such that
 $$R \ge 2\Theta\;,$$ for some strictly negative constant
 $\Theta$ (compare \eq{Thetadef}). We further assume that there exists a smooth, global solution of the inverse mean
  curvature flow  without critical points, with $u$ ranging from zero to infinity, vanishing on
  $\partial M^3$, with the level sets of $u$
$$\Sigma(s) = \{u(x)=s\}\;$$
being compact. Let $A_s$ denote the area of $\Sigma(s)$, and
define
\begin{equation}
  \label{sigdef}
  \sigma(s) = \sqrt{A_s} \int_{\Sigma(s)} ({}^2\cR_s - \frac12 H_s^2 - \frac 23
  \Theta) d^2\mu_s\;,
\end{equation}
where ${}^2\cR_s$ is the scalar curvature (equal twice the Gauss
curvature) of the metric induced on $\Sigma(s)$, $d^2\mu_s$ is the
Riemannian volume element associated to that same metric, and
$H_s$ is the mean curvature of $\Sigma(s)$. The hypothesis that
$du$ is nowhere vanishing implies that all the objects involved
are smooth in $s$. At $s=0$ we have $H_0=0$ and $A_0={A_{\pSo}}$
so that
\begin{eqnarray}
  \sigma(0)&=&
\sqrt{A_{\pSo}} \int_{\pSo} ({}^2\cR_0 -
  \frac 23 \Theta) d^2\mu_0
\nonumber \\ & = & \sqrt{A_{\pSo}}
  \left(8\pi(1-g_{\pSo}) - \frac 23 \Theta A_{\pSo}\right) \;.
\label{firstGPI}\end{eqnarray}

Generalising a formula of Hawking~\cite{SWH}, Gibbons
\cite[Equation~(17)]{GibbonsGPI} assigns to the $\Sigma(s)$
foliation a {\em total mass $\mghu$} via the formula
\begin{equation}\label{dHmn}
\mghu 
\equiv  \mbox{lim}_{
  \epsilon \to 0} \frac{\sqrt{A_{1/\epsilon}}}{32 \pi^{3/2}}
\int_{\{u=1/\epsilon\}}( {}^{2}{\cal R}_s - \frac{1}{2}H_s^{2}
-\frac{2}{3}\Theta) d^2\mu_s\; ,
\end{equation}
where $ A_{\alpha}$ is the area of the connected component under
consideration of the level set $\{u=\alpha\}$. It follows that
$$\lim_{s\to\infty}\sigma(s)= 32 \pi^{3/2}\mghu\;,$$
assuming the limit exists. The  generalisation in
\cite{GibbonsGPI} of \eq{monoton} establishes the inequality
\begin{equation}
  \label{monotonG}
\frac{  \partial \sigma}{\partial s }\ge 0\;.
\end{equation}
This  implies $\lim_{s\to\infty}\sigma(s)\ge \sigma(0)$, which
gives
\begin{equation}
\label{rds1x} 2  \mgh \ge
  (1-g_{\pSo}) \left(\frac{A_{\pSo}}{4\pi}\right)^{1/2}- \frac
  \Theta 3  \left(\frac{A_{\pSo}}{4\pi}\right)^{3/2} \;.
 \end{equation}Here  $A_{\pSo }$
is the area of $\pSo $ and $g_{\pSo }$ is the genus thereof.
 Equation~\eq{rds1x} is sharp --- the inequality there becomes
an equality for the \gK\ metrics \eq{Kot}.

The hypothesis above that $du$ has no critical points together
with our hypothesis on the geometry of the asymptotic ends forces
$\partial M^3$ to be connected. It is not entirely clear what is
the right generalisation of this inequality to the case where
several black holes occur, with one possibility being
\begin{equation}
\label{rdso} 2  \mgh \ge \sum_{i=1}^k
  \left((1-g_{\pSi}) \left(\frac{A_{\pSi}}{4\pi}\right)^{1/2}- \frac
  \Theta 3  \left(\frac{A_{\pSi}}{4\pi}\right)^{3/2}\right) \;.
 \end{equation}
Here the $\pSi$'s, $i=1,\ldots,k$, are the connected components of
$\pSo$, $A_{\pSi }$ is the area of $\pSi $, and $g_{\pSi }$ is the
genus thereof.
  This would be the inequality one would obtain from the
  Geroch--Gibbons argument if it could be carried through for $u$'s
  which are allowed to have critical points, on manifolds with
  $\pSinfty$ connected but $\pSo$  --- not connected.

As in the asymptotically flat case, the naive monotonicity
calculation of ~\cite{Geroch:extraction} breaks down at critical
level sets of $u$,
 as those do not have to be smooth submanifolds.  Nevertheless the
  existence of the appropriate function $u$ (perhaps with
 critical points) should probably follow from the results in
~\cite{HI1,HI3}. The open questions here are 1) a proof of
 monotonicity at jumps of the flow, where topology change might
 occur, and 2) the proof that the Hawking mass \eq{dHmn} exists,
 and equals the mass of the end under consideration. We also note
 that in the hyperbolic context it is  natural to consider not
 only
 boundaries $\partial M^3$ which are minimal, but also boundaries satisfying $$H = \pm 2\;.$$
 This is related to the discussion at the beginning of this
 section: if $\lambda=0$, then an apparent horizon corresponds to
 $H=0$; if $\Lambda=0$ and $\lambda=-1$, then a future apparent
 horizon corresponds to $H=2$, while a past apparent horizon
 corresponds to $H=-2$.

  Let us discuss some of the consequences of the (hypothetical) inequality \eq{rdso}.
  In the current setting there are some genus-related ambiguities
  in the definition of mass (see~\cite{ChruscielSimon} for a detailed discussion of various
  notions of mass for static asymptotically hyperbolic metrics),
  and it is convenient to introduce a mass parameter $m$
 defined as follows
\begin{equation}
  \label{masspar}
  m = \cases{ \mgh\;, & $\pSinfty=S^2\;,$ \cr \mgh\;, &
    $\pSinfty=T^2$, with the normalization $A'_\infty = -{12\pi}/{\Theta} \;,$ \cr
    \displaystyle\frac{\mgh}{|g_{\pSinfty} -1|^{3/2}}\;, &
    $g_{\pSinfty} > 1 \;.$ }
\end{equation}
Here $A'_\infty$ is the area of $\pSinfty$ in the metric
$d\Omega^2_k$ appearing in \eq{Kot}. {}For \gK\ metrics the mass
$m$ so defined coincides with the mass parameter appearing in
\eq{Kot} when $u$ is the ``radial" solution $u=u(r)$ of the
inverse mean curvature flow.

Note, first, that if all connected components of the horizon have
spherical or toroidal topology, then the lower bound \eq{rdso} is
strictly positive. For example, if  $\pSo=T^2$, and $\pSinfty=T^2$
as well we obtain
$$2m\ge -\frac\Lambda{3}\left(\frac{A_{\pSo}}{4\pi}\right)^{3/2}\;.$$
On the other hand if $\pSo=T^2$ but $g_{\pSinfty} >1$  from
Equation~\eq{rdso} one obtains
$$2m\ge -\frac\Lambda{3|g_\infty-1|} \left(\frac{A_{\pSo}}
  {4\pi}\right)^{3/2}\;.$$
Recall that in a large class of space-times\footnote{The
discussion that follows applies to all
  $(M^3,g,h)$'s that can be isometrically embedded into a
globally hyperbolic space-time $\mcM$ (with timelike conformal
boundary at infinity) in which the null convergence condition
holds; further the closure of the image of $M^3$ should be a
partial Cauchy surface in $\mcM$. Finally the intersection of the
closure of $M^3$ with $\Scri$ should be compact. The global
hyperbolicity here, and the notion of Cauchy surfaces, is
understood in the sense of manifolds with boundary,
see~\cite{GSWW} for details.} the
Galloway--Schleich--Witt--Woolgar
inequality~\cite{GSWW}\index{Galloway--Schleich--Witt--Woolgar
inequality} holds:
\begin{equation}
  \label{gsww}
  \sum_{i=1}^k{g_{\pSi}}\le g_\infty\;.
\end{equation}
It implies that if $\pSinfty$ has spherical topology, then all
connected components of the horizon must be spheres. Similarly, if
$\pSinfty$ is a torus, then all components of the horizon are
spheres, except perhaps for at most one which could be a torus. It
follows that to have a component of the horizon which has genus
higher than one we need $g_\infty >1$ as well.

When some --- or all --- connected components of the horizon have
genus higher than one, the right hand side of Equation~\eq{rdso}
might become negative. Minimising the generalised Penrose
inequality \eq{rdso} with respect to the areas of the horizons
gives the following interesting inequality
\begin{equation}
  \label{iPi1}
 \mgh  \ge -\frac{1}{3 \sqrt{-\Lambda}} \sum_i |g_{\pSi} -1|^{3/2}\;,
\end{equation}
where the sum is over those connected components $\pSi$ of $\pSo$
for which $g_{\pSi} \ge1$.  Equation~\eq{iPi1},
 together with the elementary
inequality $\sum_{i=1} ^N |\lambda_i|^{3/2} \le \left(\sum_{i=1}
^N
  |\lambda_i|\right)^{3/2}$, lead to
\begin{equation}
  \label{iPi3}
  m \ge -\frac{1}{3 \sqrt{-\Lambda}} \;.
\end{equation}

Similarly to the asymptotically flat case, the Geroch--Gibbons
argument establishing the inequality \eq{firstGPI} when a suitable
$u$ exists can also be carried through when $\pSo=\emptyset$. In
this case one still considers solutions $u$ of the differential
equation \eq{eqn:weak} associated with the inverse mean curvature
flow, however the Dirichlet condition on $u$ at $\pSo$ is replaced
by a condition on the behavior of $u$ near some chosen point
$p_0\in M^3$. If the level set of $u$ around $p_0$ approach
distance spheres centred at $p_0$ at a suitable rate, then
$\sigma(s)$ tends to zero when the $\Sigma(s)$'s shrink to $p_0$,
which together with the monotonicity of $\sigma$ leads to the
positive energy inequality:
\begin{equation}
  \label{gpet}
  \mgh \ge 0\;.
\end{equation}
It should be emphasised that the Horowitz-Myers solutions
\cite{HorowitzMyers} with negative mass show that this argument
breaks down  when $g_\infty=1$.

When $\pSinfty=S^2$ the inequality \eq{gpet}, with $\mgh $
replaced by the Hamiltonian mass (which might perhaps coincide
with $\mgh $, but this remains to be established), can be proved
by Witten type techniques~\cite{ChHerzlich,CJL}
(compare~\cite{AndDahl,GHHP,Wang,Zhang:hpet}). On the other hand
it follows from~\cite{Baum} that when $\pSinfty\ne S^2$ there
exist no asymptotically covariantly constant spinors which can be
used in the Witten argument. The Geroch--Gibbons argument has a
lot of ``ifs'' attached in this context, in particular if
$\pSinfty\ne S^2$ then some level sets of $u$ are necessarily
critical and it is not clear what happens with $\sigma$ at jumps
of topology. We note that the area of the horizons does not occur
in \eq{iPi3} which, when $g_{\pSinfty}>1$,  suggests that the
correct inequality is actually \eq{iPi3} rather than \eq{gpet},
whether or not black holes are present.

We close this section by mentioning an application of the
hyperbolic Penrose inequality to the uniqueness of static regular
black holes with a negative cosmological constant, pointed out
in~\cite{ChruscielSimon}. It is proved in that last reference that
for such connected black holes an inequality \emph{inverse} to
\eq{rds1x} holds, with equality if and only if the metric is the
one in \eq{Kot}. Hence a proof of the Penrose inequality would
imply equality in \eq{rds1x}, and subsequently a uniqueness
theorem for such black holes.

\subsection{Precise Formulations of the (full) Penrose Conjecture}
\label{SPFPC}

In the next two subsections we discuss formulations of the Penrose
Conjecture and possible applications of these statements to
defining quasi-local mass functionals with good properties and to
defining total mass in surprisingly large generality.  This
discussion is based on the third lecture~\cite{BrayCargese} given
by one of us (HB)  in Carg\`ese in the summer of 2002. Besides
discussing various formulations of the conjecture in this
subsection, we point out the value of its possible applications in
the next subsection, which greatly motivates trying to prove the
conjecture.

We begin with the question, ``Given Cauchy data, where is the
event horizon, and what lower bounds on its area can we make?''
Inequality \eq{Penrose} is the most general version of the Penrose
conjecture, but there are more ``local'' versions of it which have
the advantage of possibly being easier to prove. Recall, for
instance, that the exact location of event horizons can not be
determined from the Cauchy data $(M^3,g,h)$ without solving the
Einstein equations infinitely forward in time.  On the other hand,
apparent horizons $\Sigma$ can be computed directly from the
Cauchy data and are characterised by the equation
\begin{equation}\label{apparent}
   H_\Sigma = \mbox{tr}_\Sigma(h),
\end{equation}
that is, the mean curvature $H$ of $\Sigma$ equals the trace of
$h$ along $\Sigma$.  Note that in the $h=0$ case, this is the
assumption that $H = 0$, which is the Euler-Lagrange equation of a
surface which locally minimises area.  This leads to the first
formulation of the Penrose Conjecture, which seems to be due to
Gary Horowitz~\cite{Horowitz}:

\begin{conjecture}\label{c1}
Let $(M^3,g,h)$ be complete, asymptotically flat Cauchy data with
$\mu \ge |J|$ and an apparent horizon satisfying equation
\eq{apparent}.  Then
\begin{equation}\label{Penrose-f1}
   m \ge \sqrt{A/16\pi},
\end{equation}
where $m$ is the total mass and $A$ is the minimum area required
for a surface to enclose $\Sigma$.
\end{conjecture}
The logic is that since apparent horizons imply the existence of
an event horizon outside of it, and all surfaces enclosing
$\Sigma$ have at least area $A$, then inequality \eq{Penrose}
implies the above conjecture.

An alternative possibility would be to replace $A_e$ in
\eq{Penrose} by the area of the apparent horizon. We do not know
the answer to this, but a counterexample would not be terribly
surprising (although it would be very interesting). The point is
that the physical reasoning used by Penrose does not directly
imply that such a conjecture should be true for apparent horizons.
Hence, a counterexample to the area of the apparent horizon
conjecture would be less interesting than a counterexample to
Conjecture \ref{c1} or \ref{c2} since one of the latter
counterexamples would imply that there was actually something
wrong with Penrose's physical argument, which would be very
important to understand.

There are also good reasons to consider a second formulation of
the Penrose Conjecture, due to one of the authors (HB), for
$(M^3,g,h)$ which have more than one end.  We will choose one end
to be special, and then note that large spheres $S$ in the other
asymptotically flat ends are actually ``trapped,'' meaning that
$H_S < \mbox{tr}_S(h)$ (note that the mean curvatures of these
spheres is actually negative when the outward direction is taken
to be toward the special end and away from the other ends).  We
can conclude that these large spheres are trapped, if, for
example, the mean curvatures of these large spheres is $-2/r$ to
highest order and $|h|$ is decreasing like $1/r^2$ (or at least
faster than $1/r$). Hence, this condition also allows us to
conclude that there must be an event horizon enclosing all of the
other ends.  Thus, we conjecture

\begin{conjecture}\label{c2}
Let $(M^3,g,h)$ be complete, asymptotically flat Cauchy data with
$\mu \ge |J|$ and more than one end.  Choose one end to be
special, and then define $A$ to be the minimum area required to
enclose all of the other ends.  Then
\begin{equation}\label{Penrose-f2}
   m \ge \sqrt{A/16\pi},
\end{equation}
where $m$ is the total mass of the chosen end.
\end{conjecture}
We note that, more precisely, $A$ in the above conjecture is the
infimum of the boundary area of all smooth, open regions which
contain all of the other ends (but not the special end).
Equivalently (taking the complement), $A$ is the infimum of the
boundary area of all smooth, open regions which contain the
special end (but none of the other ends).  A smooth, compact,
area-minimising surface (possibly with multiple connected
components) always exists and has zero mean curvature.

The advantage of this second formulation is that it removes
equation \eq{apparent} and the need to define apparent horizons.
Also, preliminary thoughts by one of the authors (HB) lead him to
believe that the above two formulations are equivalent via a
reflection argument (although this still requires more
consideration).  In addition, this second formulation turns out to
be most useful in the quasi-local mass and total mass definitions
in the next subsection.

\subsection{Applications to Quasi-local Mass and Total
Mass}\label{SAqlm}\index{Quasi-local mass}\index{mass!quasi-local}

The ideas of this subsection are due to HB, and were greatly
influenced by and in some cases are simply natural extensions of
ideas due to Bartnik in~\cite{Bartnik89,BartnikTsingHua}. All of
the surfaces we are considering in this subsection are required to
be boundaries of regions which contain all of the other ends
besides the chosen one. Given such a surface $\Sigma$ in a
$(M^3,g,h)$ containing at least one asymptotically flat end, let
$I$ be the inside region (containing all of the other ends) and
$O$ be the outside region (containing a chosen end). Then we may
consider ``extensions'' of $(M^3,g,h)$ to be manifolds which
result from replacing the {\it outside} region $O$ in $M$ with any
other manifold and Cauchy data such that the resulting Cauchy data
$(\tilde{M}^3,g,h)$ is smooth, asymptotically flat, and has $\mu
\ge |J|$ everywhere (including along the surgery naturally). We
define a ``fill-in'' of $(M^3,g,h)$ to be manifolds which result
from replacing the {\it inside} region $I$ in $M$ with any other
manifold and Cauchy data such that the resulting Cauchy data
$(\tilde{M}^3,g,h)$ is smooth, asymptotically flat, and has $\mu
\ge |J|$ everywhere. Also, we say that a surface is
``outer-minimising'' if any other surface which encloses it has at
least as much area.  Note that for ``enclose'' to make sense, we
need to restrict our attention to surfaces which are the
boundaries of regions as stated at the beginning of this
paragraph. The notion of ``outer-minimising'' surfaces turns out
to be central to the following definitions.

Suppose $\Sigma$ is outer-minimising in $(M^3,g,h)$. Define the
{\em Bartnik outer mass} $m_{outer}(\Sigma)$ to be the infimum of
the total mass over all extensions of $(M^3,g,h)$ in which
$\Sigma$ remains outer-minimising.  Hence, what we are doing is
fixing $(M^3,g,h)$ inside $\Sigma$ and then seeing how small we
can make the total mass outside of $\Sigma$ without violating $\mu
\ge |J|$. Intuitively, whatever the total mass of this minimal
mass extension outside $\Sigma$ is can be interpreted as an upper
bound for the mass contributed by the energy and momentum inside
$\Sigma$.

The definition begs  the question, why do we only consider
extensions which keep $\Sigma$ outer-minimising? After all, we are
attempting to find an extension with minimal mass,  and one might
naively think that the minimal mass extensions would naturally
have this property anyway, and locally the minimal mass extensions
we defined above probably usually do (if they exist). However,
given any $\Sigma$, it is always possible to choose an extension
which shrinks to a small neck outside $\Sigma$ and then flattens
out to an arbitrarily small mass Schwarzschild metric outside the
small neck.  Hence, without some restriction to rule out
extensions with small necks, the infimum would always be zero.
Bartnik's original solution to this problem was to not allow
apparent horizons outside of $\Sigma$, and this works quite
nicely.\index{Bartnik's quasi-local mass}\index{mass!Bartnik's}
For technical reasons, however, we have chosen to preserve the
``outer-minimising'' condition on $\Sigma$, which allows us to
prove that $m_{outer}(\Sigma) \ge m_{inner}(\Sigma)$, defined in a
moment. (Thus, the definition given here is not identical to that
in~\cite{Bartnik89}, and we do not know whether or not it gives
the same number as Bartnik's original definition, although this is
a reasonable conjecture under many circumstances. We also note
that the work of Huisken and Ilmanen~\cite{HI2} described above
shows that, in the $h=0$ case, the Hawking mass of $\Sigma$ is a
lower bound for the total mass if $\Sigma$ is outer-minimising.
They also show that $m_{outer}(\Sigma) = m$ in the case that
$\Sigma$ is entirely outside the black hole of a time-symmetric
slice of the Schwarzschild metric of total mass $m$.   These
results support considering the outer-minimising condition in the
current context.)

Suppose again that $\Sigma$ is outer-minimising in $(M^3,g,h)$.
Define the {\em inner mass}\index{inner mass}\index{mass!inner}
$m_{inner}(\Sigma)$ to be the supremum of $\sqrt{A/16\pi}$ over
all fill-ins of $(M^3,g,h)$, where $A$ is the minimum area needed
to enclose all of the other ends of the fill-in besides the chosen
end. Hence, what we are doing is fixing $(M^3,g,h)$ outside
$\Sigma$ (so that $\Sigma$ automatically remains outer-minimising)
and then seeing how large we can make the area of the global
area-minimising surface (which encloses all of the other ends
other than the chosen one). Intuitively, we are trying to fill-in
$\Sigma$ with the largest possible black hole, since the event
horizon of the black hole will have to be at least $A$. If we
think of $\sqrt{A/16\pi}$ as the mass of the black hole, then the
inner mass gives a reasonable lower bound for the mass of $\Sigma$
(since there is a fill-in in which it contains a black hole of
that mass).

\begin{theorem}\label{t1}
Suppose $(M^3,g,h)$ is complete, asymptotically flat, and has $\mu
\ge |J|$.  Then Conjecture \ref{c2} implies that
\begin{equation}
   m_{outer}(\Sigma) \ge m_{inner}(\Sigma)
\end{equation}
for all $\Sigma$ which are outer-minimising.
\end{theorem}
{\it Sketch of proof:  } Consider any extension on the outside of
$\Sigma$ (which keeps $\Sigma$ outer-minimising) and any fill-in
on the inside of $\Sigma$ simultaneously and call the resulting
manifold $\bar{M}$. Since $\Sigma$ is outer-minimising, there
exists a globally area-minimising surface of $\bar{M}$ which is
enclosed by $\Sigma$ (since going outside of $\Sigma$ never
decreases area).  Thus, by Conjecture \ref{c2},
\begin{equation}\label{ggg}
   m \ge \sqrt{A/16\pi},
\end{equation}
for $\bar{M}$.  Taking the infimum on the left side and the
supremum on the right side of this inequality then proves the
theorem since the total mass $m$ is determined entirely by the
extension and the global minimum area $A$ is determined entirely
by the fill-in. \hfill\qed

\begin{theorem}\label{t2}
Suppose $(M^3,g,h)$ is complete, asymptotically flat, and has $\mu
\ge |J|$.  If $\Sigma_2$ encloses $\Sigma_1$ and both surfaces are
outer-minimising, then
\begin{equation}
   m_{inner}(\Sigma_2) \ge m_{inner}(\Sigma_1)
\end{equation}
and
\begin{equation}
   m_{outer}(\Sigma_2) \ge m_{outer}(\Sigma_1)
\end{equation}

\end{theorem}
{\it Sketch of proof:  } The first inequality is straightforward
since every fill-in inside $\Sigma_1$ is also a fill-in inside
$\Sigma_2$.  The second inequality is almost as straightforward.
It is true that any extension of $\Sigma_2$ (in which $\Sigma_2$
is still outer-minimising) is also an extension of $\Sigma_1$, but
it remains to be shown that such an extension preserves the
outer-minimising property of $\Sigma_1$.  However, this fact
follows from the fact that any surface enclosing $\Sigma_1$ which
goes outside of $\Sigma_2$ can be made to have less or equal area
by being entirely inside $\Sigma_2$ (by the outer-minimising
property of $\Sigma_2$).  But since $\Sigma_1$ was
outer-minimising in the original manifold, any surface between
$\Sigma_1$ and $\Sigma_2$ must have at least as much area as
$\Sigma_1$. \myqed

The last three theorems inspire the definition of
 the {\em quasi-local
mass} of a surface $\Sigma$ in $(M^3,g,h)$ to be the interval
\begin{equation}
   m(\Sigma) \equiv [m_{inner}(\Sigma), m_{outer}(\Sigma)] \subset R.
\end{equation}
That is, we are not defining the quasi-local mass of a surface to
be a number, but instead to be an interval in the real number
line.  Both endpoints of this interval are increasing when we move
outward to surfaces which enclose the original surface.  If
$\Sigma \subset (M^3,g,h)$ and $(M^3,g,h)$ is Schwarzschild data,
then this interval collapses to a point and equals the mass of the
Schwarzschild data (assuming Conjecture \ref{c2}).
Conversely, if the quasilocal mass interval of $\Sigma$ is a
point, then we expect that $\Sigma$ can be imbedded into a
Schwarzschild spacetime in such a way that its Bartnik data (the
metric, mean curvature vector in the normal bundle, and the
connection on the normal bundle of $\Sigma$) is preserved, which
is a nongeneric condition.  Hence, we typically expect the
quasi-local mass of a surface to be an interval of positive
length. We also expect the quasi-local mass interval to be very
close to a point in a ``quasi-Newtonian'' situation, where
$\Sigma$ is in the part of the space-time which is a perturbation
of Minkowski space, for example.  We point out that so far there
are not any surfaces for which we can prove that the quasi-local
mass is not a point. This is because there are very few instances
in which the inner and outer masses of a surface can be computed
at all.  These questions will have to wait until a better
understanding of the Penrose conjecture is found.
 This definition of quasi-local mass leads
naturally to definitions of {\em total inner mass},
$m_{inner}^{total}$, and {\em total outer mass},
$m_{outer}^{total}$, where in both cases we simply take the
supremum of inner mass and outer mass respectively over all
$\Sigma$ which are outer-minimising.

\begin{conjecture}  \label{asyflat}
If $(M^3,g,h)$ is asymptotically flat with total mass $m_{ADM}$,
then
\begin{equation}
    m_{inner}^{total} = m_{outer}^{total} = m_{ADM}.
\end{equation}
\end{conjecture}

Consider $(M^3,g,h)$ which is not assumed to have any asymptotics
but still satisfies $\mu \ge |J|$. Then we will say that $\Sigma$
(again, always assumed to be the boundary of a region in $M^3$) is
``legal'' if $\Sigma$ is outer-minimising in $(M^3,g,h)$ and there
exists an asymptotically flat extension with $\mu \ge |J|$ outside
of $\Sigma$ in which $\Sigma$ remains outer-minimising. Note that
$(M^3,g,h)$ is not assumed to have any asymptotics.  We are simply
defining the surfaces for which extensions with good asymptotics
exist, and giving these surfaces the name ``legal.''  Note also
that both $m_{inner}(\Sigma)$ and $m_{outer}(\Sigma)$ are
well-defined for legal $\Sigma$.  Thus, total inner mass and total
outer mass are well-defined as long as $(M^3,g,h)$ has at least
one legal $\Sigma$.  Finally, theorems \ref{t1} and \ref{t2} are
still true for legal surfaces even when $(M^3,g,h)$ is not assumed
to be asymptotically flat.

During the ``50 Years'' conference in Carg\`ese, Mark Aarons asked
the question, ``When are the total inner mass and the total outer
mass different?''  This is a very hard question, but it is such a
good one that it deserves some speculation.  If we define
\begin{equation}
   m_{total} \equiv [m_{inner}^{total}, m_{outer}^{total}],
\end{equation}
then total mass is very often well-defined (as long as there is at
least one legal $\Sigma$), but is not necessarily a single value.
Mark's question is then equivalent to, ``When is the total mass
single-valued and therefore well-defined as a real number?''

According to conjecture \ref{asyflat}, we are not going to find an
example of total inner mass $m_{inner}^{total} \ne
m_{outer}^{total}$, the total outer mass, in the class of
asymptotically flat manifolds. In fact, we are not aware of any
examples of $m_{inner}^{total} \ne m_{outer}^{total}$, though one
can give arguments to the effect that such situations could occur.
On the other hand we believe that in reasonable situations this
will not happen:

\begin{conjecture}
Suppose $(M^3,g,h)$ has $\mu \ge |J|$ and that there exists a
nested sequence of connected, legal surfaces $\Sigma_i = \partial
D_i \subset M$, $1 \le i < \infty$, with $\bigcup_i D_i = M$ and
$\lim |\Sigma_i| = \infty$.  Then
\begin{equation}
   m_{inner}^{total} = m_{outer}^{total} \in \R \cup \{\infty\}.
\end{equation}
\end{conjecture}
At first this conjecture seems wildly optimistic considering it is
suggesting that total mass is well-defined in the extended real
numbers practically all of the time, where the only assumptions we
are making are along the lines of saying that the noncompact end
must be ``large'' in some sense.  Note, for example, we are ruling
out cylindrical ends and certain types of cusp ends.  However, the
idea here is that most kinds of ``crazy asymptotics'' cause both
$m_{inner}^{total}$ and $m_{outer}^{total}$ to diverge to
infinity. Hence, the reason this conjecture (or one similar to it)
has a decent chance of being true is the possibility that either
$m_{inner}^{total}$ or $m_{outer}^{total}$ being finite is
actually a very restrictive situation. For example, if either the
total inner mass or the total outer mass is finite, then it might
be true that this implies that $(M^3,g,h)$ is asymptotic to data
coming from a space-like slice of a Schwarzschild space-time (in
some sense).  In this case, one would expect that both the total
inner and outer masses actually equal the mass of the
Schwarzschild space-time and therefore are equal to each other.
Certainly in the case that $(M^3,g,h)$ is precisely a slice (even
a very weird slice) of a Schwarzschild space-time, it is only
natural to point out that total mass should be well-defined.
These definitions seem to be an approach to defining total mass in
these more general settings.  However, a complete understanding of
these definitions clearly depends on making further progress
studying the Penrose Conjecture.

\bibliographystyle{amsplain}
\bibliography{../../../../../prace/references/reffile,%
../../../../../prace/references/bibl,%
../../../../../prace/references/Energy,%
../../../../../prace/references/hip_bib,%
../../../../../prace/references/netbiblio,%
../../../../../prace/references/addon,%
../../../../../prace/references/newbiblio,%
../../../../../prace/references/newbib}

\printindex

\end{document}